\documentclass[fleqn,usenatbib]{mnras}

\usepackage{newtxtext,newtxmath}

\usepackage[T1]{fontenc}

\DeclareRobustCommand{\VAN}[3]{#2}
\let\VANthebibliography\thebibliography
\def\thebibliography{\DeclareRobustCommand{\VAN}[3]{##3}\VANthebibliography}

\usepackage{graphicx}	
\usepackage{amsmath}	
\usepackage{booktabs}

\newcommand\editone[1]{\textcolor{black}{#1}}

\title[Polarimetric Tomography using Synthetic Images]{Polarimetric Tomography Applied to Synthetic Multi-Spacecraft White-Light Images: Observing Coronal Mass Ejections in 3D}

\author[D. Barnes et al.]{
David Barnes,$^{1}$\thanks{E-mail: david.barnes@stfc.ac.uk (DB)}
Erika Palmerio,$^{2}$
Tanja Amerstorfer,$^{3}$
Eleanna Asvestari,$^{4}$
Luke Barnard,$^{5}$
\newauthor{
Maike Bauer,$^{3}$
Ja\v{s}a \v{C}alogovi\'{c},$^{6}$
Greta M. Cappello,$^{7}$
Phillip Hess,$^{8}$
and Christina Kay$^{9}$
}
\\\\
$^{1}$UKRI\,--\,RAL Space, Rutherford Appleton Laboratory, Harwell Campus, Oxfordshire, UK\\
$^{2}$Predictive Science Inc., San Diego, CA, USA\\
$^{3}$Austrian Space Weather Office, GeoSphere Austria, Graz, Austria\\
$^{4}$Department of Physics, University of Helsinki, Helsinki, Finland\\
$^{5}$Department of Meteorology, University of Reading, Reading, UK\\
$^{6}$Faculty of Geodesy, University of Zagreb, Zagreb, Croatia\\
$^{7}$Institute of Physics, University of Graz, Graz, Austria\\
$^{8}$U.S.\ Naval Research Laboratory, Washington, DC, USA\\
$^{9}$The Johns Hopkins University Applied Physics Laboratory, Laurel, MD, USA
}

\date{Accepted XXX. Received YYY; in original form ZZZ}

\pubyear{\the\year{}}

\begin{document}
\label{firstpage}
\pagerange{\pageref{firstpage}--\pageref{lastpage}}
\maketitle

\begin{abstract}
A discrete tomography method has been developed that is able to reconstruct three-dimensional coronal mass ejection (CME) density structure. We test the method by producing synthetic coronagraph imagery for three events using the CORona--HELiosphere (CORHEL) model. We combine images from different numbers of observing spacecraft and we perform the method separately using polarimetric and non-polarimetric reconstructions, as a means to test their relative effectiveness. We show that increasing the number of observing spacecraft consistently reduces the mean relative absolute error (MRAE) between the simulated and reconstructed density. Furthermore, the MRAE is generally lower when using polarimetric reconstructions compared to non-polarimetric reconstructions. Methods applied to localise the CME front work well for all spacecraft configurations, and are improved when using polarimetric, over non-polarimetric, reconstructions. The presence of a CME front can be identified with an accuracy of $(72\pm9)\%$, $(70\pm8)\%$ and $(52\pm12)\%$ for CME1, CME2 and CME3 via polarimetric reconstructions using only three spacecraft at L1, L4 and L5. The radial position of the CME front can be constrained to a high level of precision when using polarimetric reconstructions using the same three spacecraft; $0.003\pm0.004$\,au, $0.004\pm0.005$\,au and $0.005\pm0.004$\,au for CME1, CME2 and CME3, respectively. We expect that at least four spacecraft are required in order to derive accurate information about 3D CME structure. We find no strong evidence of improvement when including out-of-ecliptic observers, but that their inclusion increases the volume of space within which the inversion can be performed.
\end{abstract}




\section{Introduction}

\label{sec1}
Earth-impacting CMEs can be responsible for significant geomagnetic disturbances, which can have a severe impact on both ground- and space-based technologies. Consequently, forecasting their arrival and severity is a major field of study \citep[e.g.,][]{2019Kilpua, 2019Vourlidas}. Accurate remote-sensing measurements of CMEs during their onset and propagation are understood to be key to addressing this problem \citep[e.g.,][]{2014Posner,2017Harrison}. Typically, CMEs are observed via remote sensing in the corona using coronagraph observations and, as they propagate through interplanetary space, via wide-angle heliospheric imagers. Forward modelling is often applied to such observations, whereby a pre-defined model is used to describe the CME structure and the parameters that govern the model are constrained using the available imagery \citep[e.g.,][]{2012Davies, 2016Rollett, 2018Mostl}. Forward modelling is the foremost method used to study CMEs via remote sensing because it can be applied to observations from a small number of vantage points. Indeed, many CME studies use a single observing spacecraft to constrain the model \citep[e.g.,][]{2004Yashiro,2009Gopalswamy,2009Thernisien,2017Vourlidas,2019Barnes}. The method may also be applied to noisy or low-quality data because it depends only on locating features within an image, typically the CME leading edge, or \emph{front}, \citep[e.g.,][]{2020Barnard}. In orbit at the Sun--Earth L1 point since 1995, the LASCO coronagraphs \citep{1995Brueckner} onboard the Solar and Heliospheric Observatory (SOHO) provide such observations, as do the remote sensing camera suites onboard the two Solar Terrestrial Relations Observatory (STEREO) spacecraft (COR-1 and -2 coronagraphs: \citealp{2008Howard} and HI-1 and -2 heliospheric imagers: \citealp{2009Eyles}), launched in 2006. Consequently, many forward modelling techniques have been developed that use images from two \citep[e.g.,][]{2009deKoning,2010Byrne,2013Davies} or three \citep[e.g.][]{2009Thernisien} spacecraft to better constrain CME parameters. Such models are widely used in scientific studies of CMEs \citep[e.g.,][]{2013Shen,2019Pluta,2020bBarnes,2024Gandhi} as well as in practical space weather forecasting situations \citep[e.g.,][]{2013Colaninno,2017Kay,2023Palmerio,2024Laker}.

Inverse modelling is an approach whereby white-light images are treated as two-dimensional projections of the Thomson-scattered light from a three-dimensional plasma distribution. Inversion of these images, taken from multiple vantage points, is purely mathematical and allows the structure to be constrained. The first such methods made use of solar rotation to observe quasi-static solar wind structures from multiple vantage points using a single spacecraft (e.g. \citealp{1977Kastner} with OSO-7 and \citealp{1979Altschuler} with Skylab), a technique known as \emph{rotational tomography}. This technique is used extensively in more modern studies \citep[e.g.,][]{1997Zidowitz,1999Zidowitz,2002Frazin,2010Morgan,2013aDePatoul,2015Morgan,2019Morgan}. Furthermore, the nature of Thomson scattering in the heliosphere means that polarised brightness (pb) measurements provide more information with which to constrain density structures \citep{2009Howard,2016aDeForest}, known as \emph{polarimetric reconstructions} \citep[e.g.,][]{2004Moran,2005Dere}.

Multi-spacecraft tomography was first applied by \cite{1995Jackson} to observations from the zodiacal light photometers on-board the two HELIOS spacecraft and the Solwind coronagraph. Since the launch of STEREO, many attempts have been made using the combined vantage points of STEREO and SOHO to perform tomographic inversions using two or three spacecraft. \cite{2009Antunes} achieve this through a combination of forward and inverse modelling using the Graduated Cylindrical Shell \citep[GCS;][]{2006Thernisien,2009Thernisien} model; \cite{2009Frazin}, too, combine inverse modelling with a separate CME model and \cite{2012Frazin} complement the inversion with magnetohydrodynamic (MHD) modelling. Efforts by \cite{1997Jackson,1998Jackson,1998Kojima,1998Asai} led to the development of a method that uses the solar rotation technique and radio interplanetary scintillation (IPS) measurements to constrain a heliospheric density model. \cite{1998Jackson} also employ physical principles -- radial solar wind outflow and conservation of mass and mass flux -- to optimise the solution. Further developments of this technique are well summarised in \cite{2006Jackson} and are a proven method of driving space weather forecasting models \citep{2015Yu,2015Jackson}. It is apparent that these multi-spacecraft methods must make use of CME modelling, as well as physical assumptions, in order to constrain CME structure, and it is consistently noted in the literature that two or three observing spacecraft are too few to perform a classical inversion of CME white-light observations \citep[e.g.,][]{2009Frazin,2020aBarnes}. Indeed, a study by \cite{1994Davila} shows, using simple mathematical modelling, that at least four spacecraft would be required to resolve density structures in the corona via a pure inversion of the data. 

Inverse modelling is a form of photometric image analysis and, as such, requires a high standard of image processing, particularly at large distances from the Sun, in order to separate the Thomson-scattered K-corona from F-coronal background, stars, planets, and image artefacts. Typically, forward modelling often uses the running-difference of an image sequence \citep[e.g.,][]{1999Sheeley,2008Sheeley}, in order to highlight moving structures. Simply identifying CME features requires signal separation with a precision of $10^{-2}$, whilst thorough photometric analysis can require $10^{-4}$ at 45$^\circ$ from the Sun \citep{2016aDeForest}. The latter is indeed achieved using in-depth image processing by \cite{2011DeForest}, a technique that is able to resolve solar wind substructure \citep{2012Howard,2015DeForest} and CME substructure \citep{2013bDeForest,2014Howard} well into the field of view (FOV) of HI-2. Advanced imaging systems capable of polarised brightness measurements out to large elongation angles onboard PUNCH \citep{2026DeForest}, which recently launched on 11 March 2025, are expected to lead to further advances in both image processing and photometric analysis techniques \citep{2013aDeForest,2013bDeForest}.

With the continued operation of SOHO and STEREO-A, and the more recent launches of Solar Orbiter \citep{2020Muller,2021GarciaM} and Parker Solar Probe \citep{2016Fox}, multi-spacecraft CME studies are now commonplace \citep[e.g.,][]{2021Davies,2022Mostl,2023Niembro,2023Liberatore}. Newly launched and upcoming missions such as SWFO-L1 \citep{2017Kraft} and Vigil, respectively, recent ambitious proposals such as the Solar Ring mission \citep{2020Wang} and contemplation of future missions to polar \citep{2023Hassler} and L4 \citep{2021Posner} orbits mean that it is a most opportune time to investigate the potential of multi-spacecraft remote-sensing observations with respect to CME analysis. Furthermore, continuing improvements in image processing mean that we possess an increasing ability to perform accurate photometric analysis on these observations. 

Consequently, our aim is to establish the effectiveness of the tomographic inversion method applied to white-light images using multiple spacecraft, and we achieve this by means of synthetic imagery. Specifically, we refer to a three-dimensional, discrete inversion of the data using no forward modelling nor physical assumptions beyond those of Thomson scattering. We achieve the synthetic images by employing state-of-the-art MHD simulations of the solar corona using the Magnetohydrodynamic Algorithm outside a Sphere \citep[MAS;][]{1999Mikic, 2018Mikic} code, which provides time-dependent, three-dimensional plasma parameters, including coronal density. From these densities we are then able to create sequences of synthetic coronagraph images, including polarised brightness measurements. This is performed for a fleet of spacecraft, such that various combinations of virtual probes can be used to perform tomography on the synthetic imagery, with the goal of establishing the minimum requirements for such a technique to succeed. Furthermore, we are able to investigate how the number of observing spacecraft influences the solution, how well the technique is augmented using polarised brightness measurements and the optimal orbital configuration, including out-of-ecliptic observers. 

\section{Methods}

In this section, we first briefly describe the MHD model employed in this work, the configuration of the synthetic spacecraft considered throughout our investigation, and the synthetic CME case studies produced for our analysis (Section~\ref{subsec:sims}). Then, we provide an overview of the tomographic inversion method employed and the image processing performed on the synthetic data (Section~\ref{subsec:tomo}), followed by a description of the validation methodology that we consider (Section~\ref{subsec:validation}).

\subsection{MHD Simulations with CORHEL-CME} \label{subsec:sims}

\subsubsection{The MAS Code and CORHEL Model}

We employ the MAS code to model the three synthetic CME events explored throughout this work. To set up and run the eruption and evolution of the CMEs we use the CORHEL-CME \citep{2024Linker} suite, a graphical-user-interface-based system derived from the CORona--HELiosphere \citep[CORHEL;][]{2012aRiley} model that simulates CME evolution from the Sun to Earth. The MAS code integrates the time-dependent resistive thermodynamic MHD equations on a non-uniform spherical mesh. By employing photospheric magnetic field observations as boundary conditions on the radial magnetic field, MAS allows modelling of the solar corona \citep[e.g.,][]{2023Lionello, 2025Downs}, the inner heliosphere \citep[e.g.,][]{2011Riley, 2012bRiley}, as well as transient events within them \citep[e.g.,][]{2013Lionello, 2018Torok}. CORHEL operates within two main domains: the corona (COR), which is usually set in the range 1--30\,$R_{\odot}$, and the heliosphere (HEL), usually spanning the interval 28--230\,$R_{\odot}$. 

To model the three synthetic CMEs, we first use a photospheric magnetic field map to generate a relaxed, steady-state coronal and heliospheric background solution. Subsequently, we insert a flux rope at the polarity inversion line (PIL) of interest using the Regularized Biot--Savart Law \citep[RBSL;][]{2018Titov} description. Since in this work our aim is to investigate the post-eruptive structure and evolution of CMEs in the corona, in all cases we insert out-of-equilibrium flux ropes that are expected to erupt at the very beginning of each simulation. We drive each run for both COR and HEL for a temporal duration of six days, well after any of the three CMEs has reached 1~au. Nevertheless, in this study, we focus exclusively on results within the COR domain, whilst deferring a detailed analysis of the three CMEs in the inner heliosphere to future work.

\subsubsection{Synthetic Images and Spacecraft Configurations} \label{subsubsec:sc}

After the three CME runs have been performed, we generate synthetic white-light imagery within the coronal domain from different viewpoints realised via the placement of virtual probes at 1~au through the heliospheric domain. To obtain the synthetic observables of interest, we integrate the time-dependent COR density data cubes within the range 2--30\,$R_{\odot}$ and produce total brightness \citep[tb; Equation~29 in][]{2009Howard} as well as polarised brightness \citep[pb; Equation~24 in][]{2009Howard} coronagraph-like images. We employ a field-of-view (FOV) extending to $8^\circ$, which emulates the LASCO/C3 coronagraph. For each event and for each observing location, we produce imagery at 5-minute cadence for the first 12~hours of simulation, to ensure that all CMEs have fully traversed the COR domain. We remark that our synthetic images contain the K-corona only (as the F-corona is not present within the CORHEL modelling suite), and thus are not affected by the same background subtraction issues as in real data.

\begin{figure}
    \centering
    \includegraphics[width=0.45\textwidth]{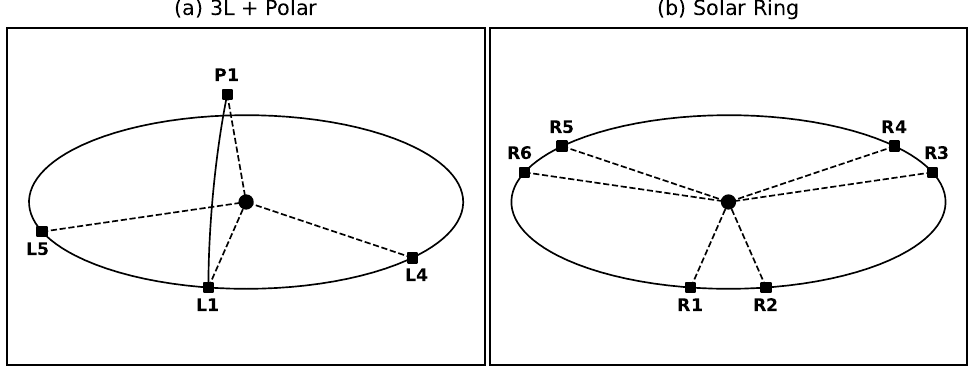}
    \caption{Sketch illustrating the placement of the synthetic spacecraft placed through the MHD simulation data cubes and two of the configurations considered in this work -- see Table~\ref{tab:orbit_configs} for a full list. (a) The `4sc' configuration, with observers placed at L1, L4, L5, and a polar orbit P1. (b) The `6sc' configuration, featuring three pairs of spacecraft in a ring formation at 1~au from the Sun.}
    \label{fig:sc_cartoon}
\end{figure}

Within the study, we employ five different combinations of observing spacecraft to perform the tomographic inversion for each of the three case studies. We focus on logistically feasible orbits, such as the L1, L4, and L5 points, groups of spacecraft in the ecliptic, and an out-of-ecliptic observer. Figure \ref{fig:sc_cartoon} shows two such combinations. Panel (a) shows four spacecraft at the L1, L4, L5 positions, and P1 in a polar orbit $60^\circ$ above the ecliptic. Panel (b) shows six spacecraft in a ring throughout the ecliptic, formed of three equally spaced pairs, where the spacing within each pair is $20^\circ$, akin to the originally proposed Solar Ring mission \citep{2020Wang}. Within the simulations, R1 is placed at the L1 point, meaning that images from L1 and R1 are identical. We therefore have a total of nine spacecraft, from which we create the five observing configurations listed in Table~\ref{tab:orbit_configs}. This includes the two configurations shown in Figure~\ref{fig:sc_cartoon}, plus a three-spacecraft combination consisting of L1, L4 and L5, a three-spacecraft combination consisting of L1, R3, and R5 and, lastly, a seven-spacecraft fleet consisting of the six solar ring spacecraft plus the out-of-ecliptic P1.

\begin{table}
    \centering
    \begin{tabular}{c|c|c|c|c|c|c|c|c|c}
         label    & L1 & L4 & L5 & P1 & R2 & R3 & R4 & R5 & R6  \\
         \toprule
         3sc      & \checkmark & \checkmark & \checkmark &            &            &            &            &            & \\
         3sc ring & \checkmark &            &            &            &            & \checkmark &            & \checkmark & \\
         4sc      & \checkmark & \checkmark & \checkmark & \checkmark &            &            &            &            & \\
         6sc & \checkmark &            &            &            & \checkmark & \checkmark & \checkmark & \checkmark & \checkmark \\
         7sc      & \checkmark &            &            & \checkmark & \checkmark & \checkmark & \checkmark & \checkmark & \checkmark \\
    \end{tabular}
    \caption{List of synthetic spacecraft orbital configurations used in the study. \emph{4sc} and \emph{6sc} refer to panels (a) and (b) of Figure \ref{fig:sc_cartoon}, respectively.}
    \label{tab:orbit_configs}
\end{table}

\subsubsection{CME Case Studies}

\begin{figure*}
    \centering
    \includegraphics[width=0.99\textwidth]{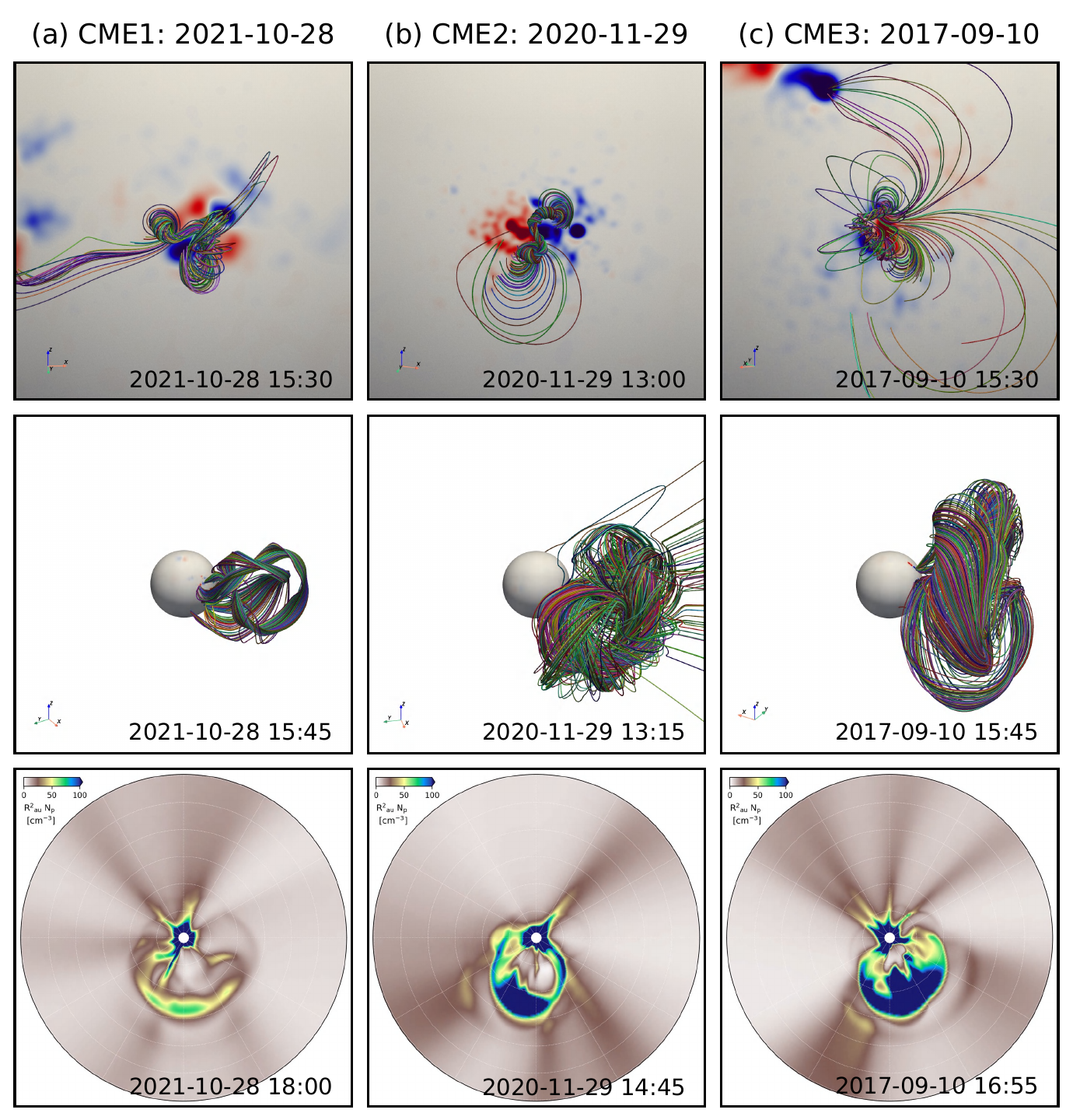}
    \caption{Overview of the three CMEs simulated with CORHEL-CME, showing (a) CME1, (b) CME2, and (c) CME3. For each event, the top panel shows selected field lines of the pre-eruptive (or at $t=0$ in the simulation) RBSL flux rope over the source AR on the photosphere, the middle panel displays CME field lines 15~minutes after the eruption, and the bottom panel features a slice of the CME scaled density along the solar equatorial plane at the time when the front reaches approximately 15~$R_{\odot}$ (the plot covers the full COR domain, i.e.\ up to 30~$R_{\odot}$).}
    \label{fig:mhd_sims}
\end{figure*}

The three CMEs modelled and analysed in this work are all inspired by real events, meaning that their source regions and pre-eruptive configurations are based on observed CMEs, but no further attempt was made to reproduce their coronal and/or heliospheric evolution. All the photospheric synoptic maps employed as boundary conditions use data from the Helioseismic and Magnetic Imager \citep[HMI;][]{2012Scherrer} onboard the Solar Dynamics Observatory \citep[SDO;][]{2012Pesnell} -- specifically, we join two successive Carrington maps so that each heliolongitude in the final magnetogram features data collected as close as possible to the selected CME eruption time. When inserting the RBSL flux rope at the active region (AR) of interest, we base our initial parameters on the corresponding ``real'' events -- in particular, we select the locations of the flux rope footpoints as to reproduce the observed flux rope type \citep[i.e., its axial magnetic field orientation and helicity sign; see][and references therein]{2017Palmerio} and we vary the initial axial current to loosely emulate the observed CME speed in the corona. For each event, we place L1 approximately in front of the CME source longitude (regardless of the real position of Earth) and distribute the remaining virtual spacecraft accordingly (see Section~\ref{subsubsec:sc}). An overview of the three events considered in this work is shown in Figure~\ref{fig:mhd_sims}, and in the remainder of this section we provide a brief overview of the three CMEs.

The first event (CME1) is inspired by the 2021 October 28 CME, which erupted around 15:30~UT from the vicinity of the central meridian as seen from Earth \citep[e.g.,][]{2023Chikunova, 2024Kouloumvakos}. To model this event, we place an RBSL flux rope in AR~12887, designing a right-handed structure with its axis roughly parallel to the solar equatorial plane (Figure~\ref{fig:mhd_sims}(a), top). The resulting CME in the corona is the smallest in our data set (Figure~\ref{fig:mhd_sims}(a), middle) and is characterised by relatively modest speeds, moving at approximately 1000~km~s$^{-1}$ between the altitudes of 5 and 10~$R_{\odot}$ (based on visual inspection of MHD quantity slices on equatorial and meridional planes such as the one displayed in Figure~\ref{fig:mhd_sims}(a), bottom). 

The second event (CME2) is inspired by the 2020 November 29 CME, which erupted around 13:00~UT from the vicinity of the eastern limb as seen from Earth \citep[e.g.,][]{2022Nieves-Chinchilla, 2022Palmerio}. To model this event, we place an RBSL flux rope in AR~12790 (which was still an unnamed region at the time of the event), designing a right-handed structure with its axis roughly perpendicular to the solar equatorial plane (Figure~\ref{fig:mhd_sims}(b), top). The resulting CME in the corona is larger (Figure~\ref{fig:mhd_sims}(b), middle) and faster than the previous event, with speeds between 5--10~$R_{\odot}$ of approximately 1500~km~s$^{-1}$.

The third event (CME3) is inspired by the 2017 September 10 CME, which erupted around 15:30~UT from the vicinity of the western limb as seen from Earth \citep[e.g.,][]{2018Veronig, 2020Scolini}. To model this event, we place an RBSL flux rope in AR~12673, designing a left-handed structure with a complex V-shaped axial configuration (Figure~\ref{fig:mhd_sims}(c), top). The resulting CME in the corona is the largest and fastest in our data set, its speed being approximately 2000~km~s$^{-1}$ as it propagates between 5 and 10~$R_{\odot}$.

Thus, the three CMEs modelled and analysed in this work span a wide range of speeds, sizes, and axial orientations: CME1 is characterised by a low speed, a small size, and an overall east--west axial orientation; CME2 is characterised by an intermediate speed, an intermediate size, and an overall north--south axial orientation, and CME3 is characterised by a high speed, a large size, and a complex, compound axial orientation.

\subsection{Tomographic Inversion} \label{subsec:tomo}

The range of methods in solar and heliospheric image analysis that are described using the term \emph{tomography} is broad. For clarity, we reiterate that the technique performed here is discrete tomography. The volume encompassing the solar corona is divided into a spherical, heliocentric grid, which, in this study, has a resolution of $\frac{1}{200}$\,au $\times3^\circ\times3^\circ$. Each grid cell contains an unknown density value and the radiance measured by every image pixel is used to constrain this density distribution. This is realised by defining the equation

\begin{equation}
    \mathbf{y}=\mathbf{Hx},
    \label{eq:inverse}
\end{equation}

\noindent
where \textbf{y} is an array containing the observed radiance values from every spacecraft, \textbf{x} is an array containing the unknown density values over the grid, and \textbf{H} is a physical operator, defined by the laws of Thomson scattering, that relates the discrete values of \textbf{y} and \textbf{x}. Equation \eqref{eq:inverse} is then an inverse relation that can be solved for \textbf{x}, which is achieved by means of an iterative convergence algorithm. Before solving the inverse equation, we first apply some processing to the images in order to separate the CME structure from the background density and we apply some regularisation to the equation before calculating the density, the details of which are described in the following sections.

\subsubsection{Image Processing}

\begin{figure*}
    \centering
    \includegraphics[width=0.99\textwidth]{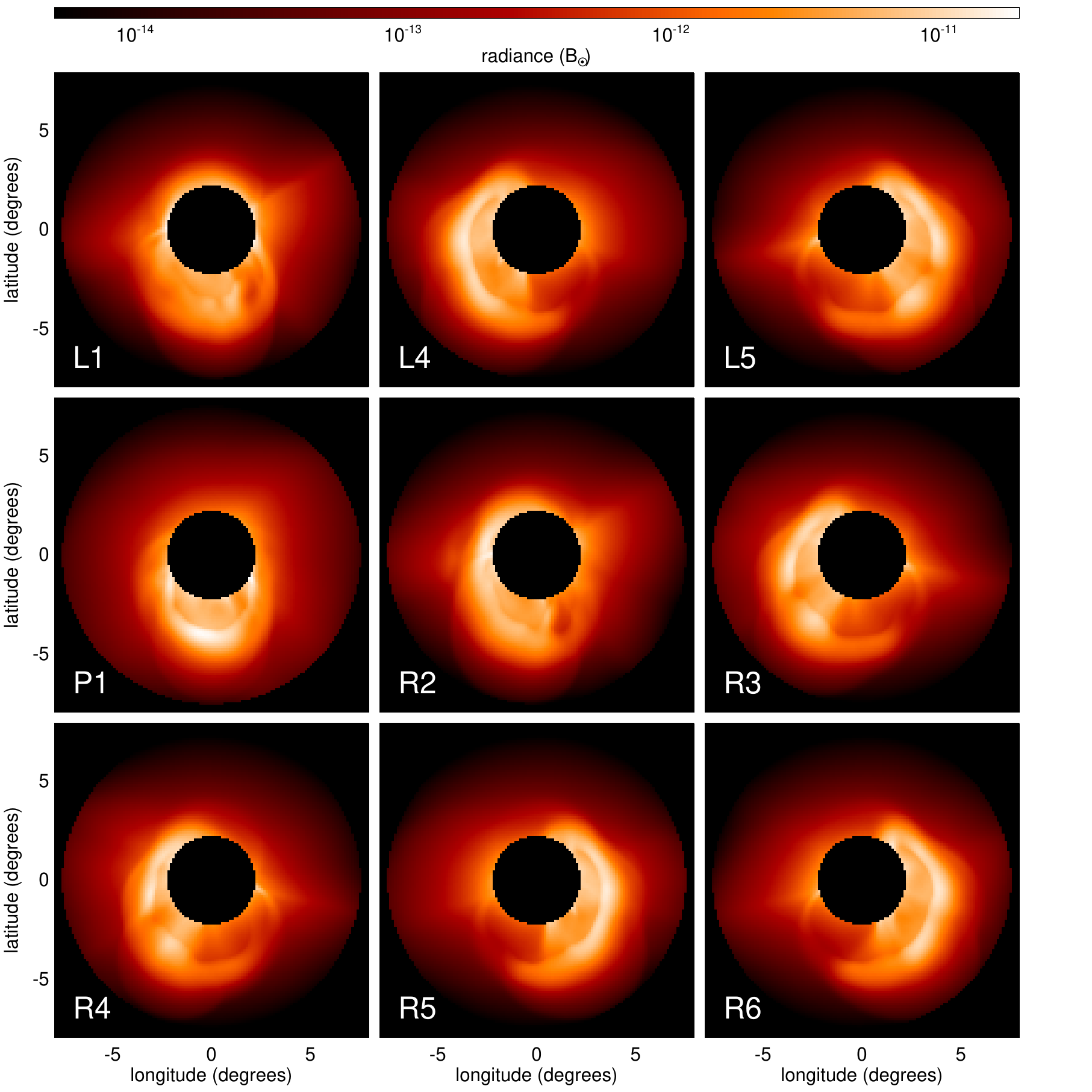}
    \caption{Synthetic total-brightness coronagraph images of the 29/11/2020 event (CME2) at $t=28$ for each of the nine spacecraft. Each image has been processed to subtract the background and reduced to $128\times128$ pixels with a 4$R_\odot$ occulter applied to obscure pixels near the Sun.}
    \label{fig:sim_image_9}
\end{figure*}

The main image processing techniques typically applied to real white-light images are designed to separate the K-corona, F-corona, and background starlight, usually to study the former. The synthetic coronagraph images used in this study contain only K-corona; however, we still perform some processing to separate the CME structure from the background density in order to perform the inversion. To process an image at a given time-step, we take the current image and the six previous images in the sequence. The minimum of the seven values in a given pixel is calculated and multiplied by 0.75, which is assumed to be the background. This is then subtracted from the original image to reveal the CME structure. The factor of 0.75 is chosen because it is sufficient to reveal the CME structure, whilst also resulting in realistic measurements of density in the inversion. Subtracting more of the background makes the resulting densities much lower than in the simulations, whilst subtracting less inhibits the inversion method because the background signal dominates the images. We perform the same processing on the associated polarised brightness images, with an extra stipulation that the degree of polarisation (the ratio of polarised- to total-brightness) cannot exceed 1.

Using the full $2048\times2048$ resolution images in our analysis would make solving Equation \eqref{eq:inverse} intractable, so we reduce the image resolution to $128\times128$ pixels by averaging the radiance values in $16\times16$ pixel bins. Furthermore, we occult pixels within $2.16^\circ$ (4$R_\odot$) of the image centre, as a means to exclude structures close to the Sun that dominate the image brightness. We also occult pixels outside $7.60^\circ$ degrees, to remove adverse polarisation effects that occur near the edge of the FOV due to the nature of the simulations. Figure \ref{fig:sim_image_9} shows an example of the processed $128\times128$ pixel images for each of the 9 spacecraft configurations taken at $t=28$ for CME2. We note that the pairs of spacecraft L4+R5 and L5+R3 are separated by $180^\circ$ in longitude and they effectively observe mirror images of the CME. As such we never include both spacecraft from either pair in the same configuration.

Finally, we multiply all pixels by a weighting factor given by Equation \eqref{eq:weight}, similar to that of \cite{2009Kramar}, to account for the large dynamic range in the data. This effectively flattens the distribution of observed radiance values. However, it causes the solving algorithm to slightly disfavour the very dimmest pixels, meaning that they are less well constrained.

\begin{equation}
  w_j = \frac{1}{y_j+\operatorname{min}(\textbf{y})}
  \label{eq:weight}
\end{equation}

\subsubsection{Inversion Method}
The tomographic inversion method employed in this study is principally the same as that detailed in \cite{2020aBarnes}, which the reader is encouraged to use for reference. We differ only in that the analysis has been extended to three-dimensions and that it employs a full integration over the photosphere. For the sake of brevity, a summary of the method is presented here. The radiance measured by an image pixel consists of three integrals \citep{2009Howard}: an integration over the photosphere, an integration over the point-spread-function of the pixel and an integration along the line-of-sight (LOS). The first of these integrals is accounted for using the Van de Hulst coefficients to simplify the equations, and the second is ignored by assuming that electron density does not vary across the perpendicular field of a single pixel. The third integral is utilised in order to perform the tomographic inversion. For a given pixel LOS, its path through the pre-defined grid is calculated using the method described in \cite{2009Tappin}. The LOS integral is then approximated as a sum of finite elements, each of which is the contribution from an individual grid cell to the measured radiance, given by the following equation

\begin{multline}
  H_{ij}=\frac{\pi\sigma_{e}}{2}\Bigl(2[(1-uC_i)+uD_i]- \\ \sin^{2}{(\phi_i+\epsilon_j)}[(1-uA_i)+uB_i]\Bigr)dz_i,
  \label{eq:elements_tb}
\end{multline}

\noindent
where $\sigma_e$ is the electron differential cross section for perpendicular scattering; $u$ is the solar limb darkening coefficient; $A$, $B$, $C$ and $D$ are the Van de Hulst coefficients, which are listed in \cite{2009Howard}; $dz$ is the distance through a grid cell along the pixel LOS; $\phi$ is the observer\,--\,Sun\,--\,electron angle and $\epsilon$ is the pixel elongation angle. The subscripts $i$ and $j$ refer to the grid cell and the image pixel, respectively, and the value of $H_{ij}$ is otherwise zero if the grid cell $i$ is not crossed by pixel $j$. We use this to calculate the elements of \textbf{H} in Equation~\eqref{eq:inverse}, which can be understood as a system of linear equations that approximate the LOS integral for every image pixel, as a function of the unknown density over the grid. A grid cell is included in the solution volume only if it is observed by at least two spacecraft. When including polarised brightness observations, every pixel then has two values: the total brightness $y_{tb}$, corresponding to the original Equation~\eqref{eq:elements_tb}, and polarised brightness, $y_{pb}$, corresponding to extra elements given by

\begin{equation}
  H_{ij}=\frac{\pi\sigma_{e}}{2}\left(\sin^{2}{(\phi_i+\epsilon_j)}[(1-uA_i)+uB_i]\right)dz_i,
  \label{eq:elements_pb}
\end{equation}

\noindent
which doubles the size of the inverse equation \eqref{eq:inverse}. Equation \eqref{eq:elements_pb} is equal to the right-hand term in Equation \eqref{eq:elements_tb}, which is a function of the scattering angle, $\chi=\pi-(\phi+\epsilon)$, whilst the left-hand term is not. Polarised brightness measurements should therefore afford extra information with which to constrain the density distribution along a given LOS. Within the analysis, we refer to the \emph{degree of polarisation}, which is defined as the ratio of polarised- to total-brightness, $\mathbf{p}=\mathbf{y}_{pb}/\mathbf{y}_{tb}$. 

Regularisation is also applied in order to ensure some smoothness in the resulting density by adding extra terms to Equation \eqref{eq:inverse}, which are a discrete approximation of the respective gradient operators in each of the radial, azimuthal, and polar dimensions, as in \cite{2020aBarnes}. Likewise, we use the same solving algorithm -- the BiConjugate Gradient Stabilised algorithm \citep{1992VanDerVorst} -- because it can be applied to sparse, non-symmetric matrices and because it converges quickly. The algorithm starts with an initial guess of the density distribution, $\mathbf{x}^{(0)}$, and determines an updated value of $\mathbf{x}^{(i)}$ after each iteration, which is intended to decrease the residual, $|\mathbf{y}-\mathbf{Hx}^{(i-1)}|$. By minimising the residual after a sufficient number of iterations, we arrive at a solution, $\mathbf{{x}^\prime}$, that satisfies the observations in $\mathbf{y}$ to an adequate level of accuracy. We choose the value of $\mathbf{x}^{(0)}$ to be $1.0\,$\,electrons\,cc$^{-1}r^{2}$ for the first time-step, because it is a sufficient estimate with which to initialise the solving algorithm. Successive time-steps are then initialised using the solution from the prior time-step.

\subsection{Validation Methodology} \label{subsec:validation}
We will focus on three approaches as a means to quantify the fidelity of our density reconstructions for each CME and for each spacecraft configuration. Firstly, we assess how well the convergence algorithm performs; secondly, we evaluate how well the density solutions compare to the original simulations; and, finally, we ascertain how well we are able to establish the location of the CME within the solution volume. The first method is performed by creating images from our density solution, $\mathbf{{x}^\prime}$, using Equation \eqref{eq:inverse}, i.e. $\mathbf{y^\prime=Hx^\prime}$, and comparing these with the original images, $\mathbf{y}$, that were used to perform the inversion. Taking the difference of the two gives a measure of the residual and enables us to quantify the performance of the convergence algorithm. Secondly, we compare the densities derived from tomographic inversion, $\mathbf{x^\prime}$, to the original \emph{`true'} density, $\mathbf{x}$, from the MAS simulations. Because the MAS calculations are performed over a much finer grid than the tomography results, we re-bin the MAS densities on a $\frac{1}{200}\,au\times3^\circ\times3^\circ$ grid, equivalent to that used in the tomography. This is achieved by taking the nearest MAS grid cell, based on the central position, to each tomography grid cell. In order to account for the background subtraction performed on the images, we also subtract the background from the MAS densities by subtracting the density at $t_{0}$ (prior to CME eruption) from each successive time step. Finally, we apply analysis in order to locate the CME front within the 3D density distribution. This is accomplished by identifying grid cells that lie above a certain threshold value in the tomographic reconstructions, and these grid cells are then assumed to lie within the CME. For a given latitude and longitude, the most advanced grid cell lying within the CME is determined to be the CME front. The threshold is taken to be 25\% of the maximum density in the 3D distribution at a given time-step. Again we apply this analysis to densities from both the tomographic inversions and the original MAS simulations, using the same threshold value, as a means of comparison.

\section{Results and Discussion}
\subsection{Accuracy of Solving Algorithm}

\begin{figure*}
    \centering
    \includegraphics[width=0.95\textwidth]{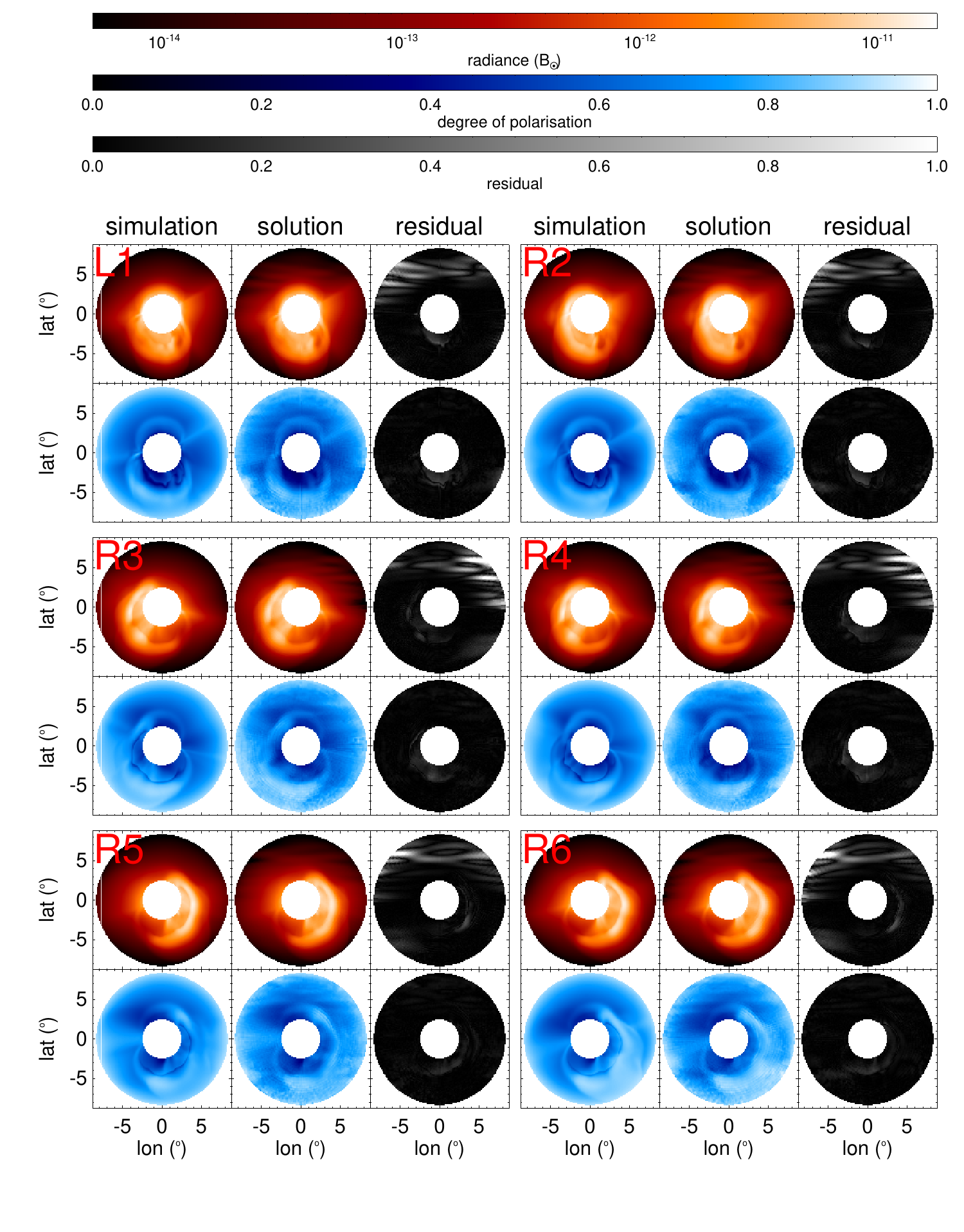}
    \caption{Processed images and the corresponding solutions and residuals from tomography applied to the 29/11/2020 event (CME 2) at $t=28$ using six observing spacecraft. Each block of six images shows \textbf{(top left)} processed image used as input to tomography, \textbf{(top middle)} image produced from tomographic density solution, \textbf{(top right)} relative absolute error between the simulation and solution images, \textbf{(bottom left)} degree of polarisation in the simulation images, \textbf{(bottom middle)} degree of polarisation from the tomographic density solution, \textbf{(bottom right)} relative absolute error between the simulation and solution degree of polarisation.}
    \label{fig:image_residuals}
\end{figure*}

\noindent
Figure \ref{fig:image_residuals} shows an example of the original ($\mathbf{y}$) and solution ($\mathbf{y^\prime}$) images, and their residual, which we display as the relative absolute difference (RAE) between the two, $|\mathbf{y-y^\prime}|/\mathbf{y}$. The values shown represent the 6-spacecraft (L1, R2\,--\,R6) polarised-brightness solution applied to CME2 at $t=28$. Each block of six panels corresponds to one of the six spacecraft, labelled in the top-left, where the total brightness values are shown in red (top row), and the degree of polarisation shown in blue (bottom row). In each block, the top-left panel corresponds to the original images produced from MAS ($\mathbf{y}_{tb}$, labelled \emph{simulation}), the top-middle panel corresponds to the images derived from the tomographic inversion ($\mathbf{y^\prime}_{tb}$, labelled \emph{solution}) and the top-right panel shows their residuals ($|\mathbf{y}_{tb}-\mathbf{y^\prime}_{tb}|/\mathbf{y}_{tb}$). Likewise, the bottom-left shows the degree of polarisation in the simulation, $\mathbf{p}$, the bottom-middle shows the degree of polarisation in the solution, $\mathbf{p^\prime}$, and the bottom-right shows their residual, $|\mathbf{p}-\mathbf{p^\prime}|/\mathbf{p}$. In these cases the mean relative absolute error (MRAE) for total-brightness measurements ranges from 0.38 (R4) to 0.39 (R6), showing good agreement between simulation and solution, which can be seen by qualitative inspection of the images. It is also apparent that, in all cases, the regions of the image where the RAE is greatest typically lie away from the main CME structure, corresponding to the dimmest pixels. This is attributed to the weighting factor (Equation \eqref{eq:weight}), which causes the solving algorithm to constrain the dimmest image pixels less accurately. We also note that the image processing and regularisation applied to the equation before solving means that we are inhibiting the ability of the solving algorithm to arrive at a result that precisely reproduces the original data. Regarding the degree of polarisation measurements, the MRAE values for all six spacecraft ranges from 0.028 (R5) to 0.032 (L1), showing that the solution accurately reproduces the polarimetric properties of the original simulation.

\begin{figure*}
    \centering
    \includegraphics[width=0.95\textwidth]{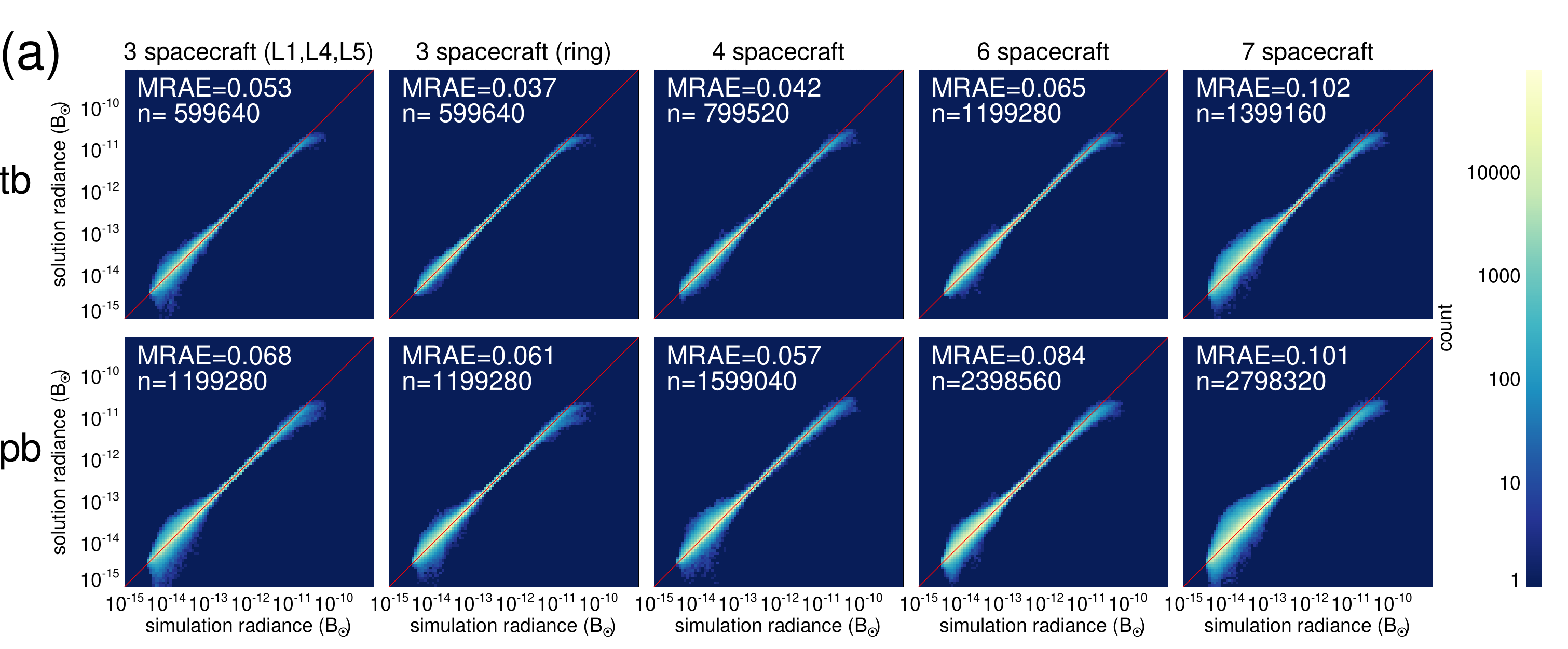}
    \includegraphics[width=0.95\textwidth]{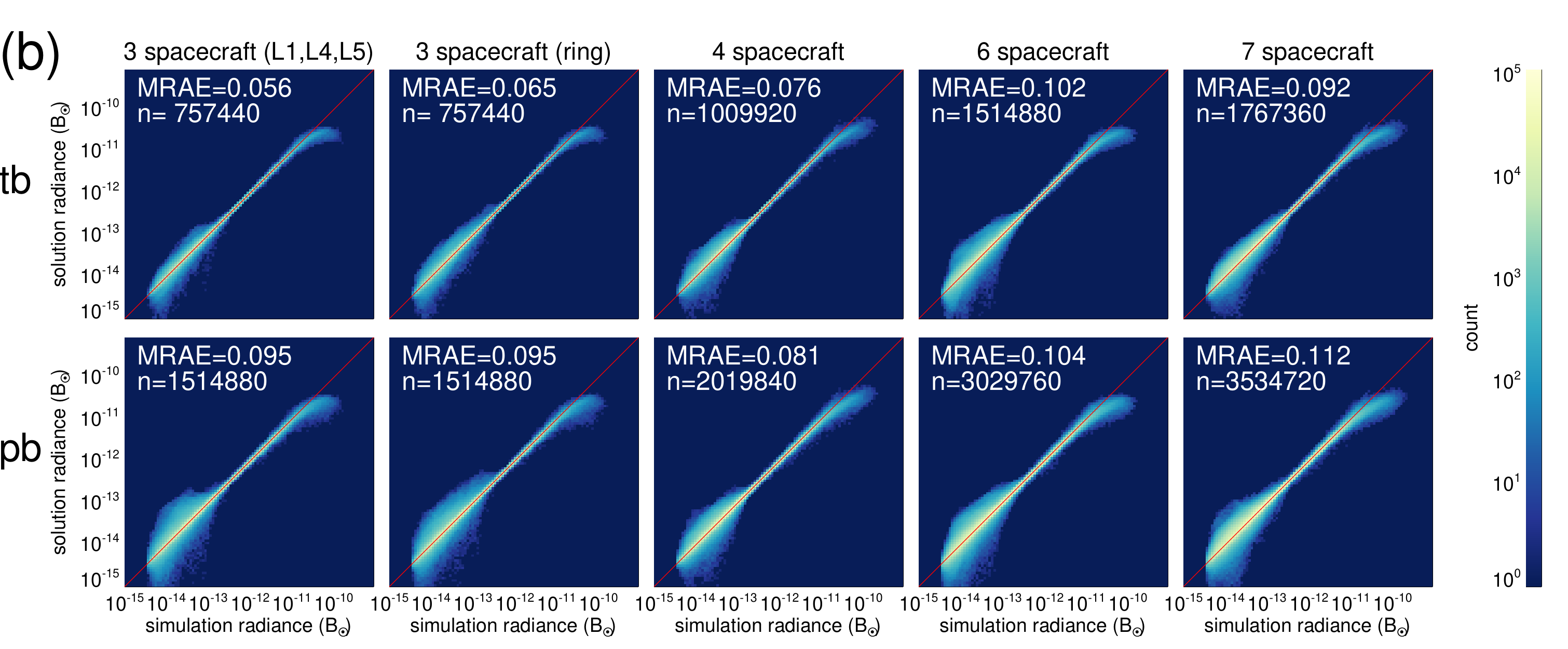}
    \includegraphics[width=0.95\textwidth]{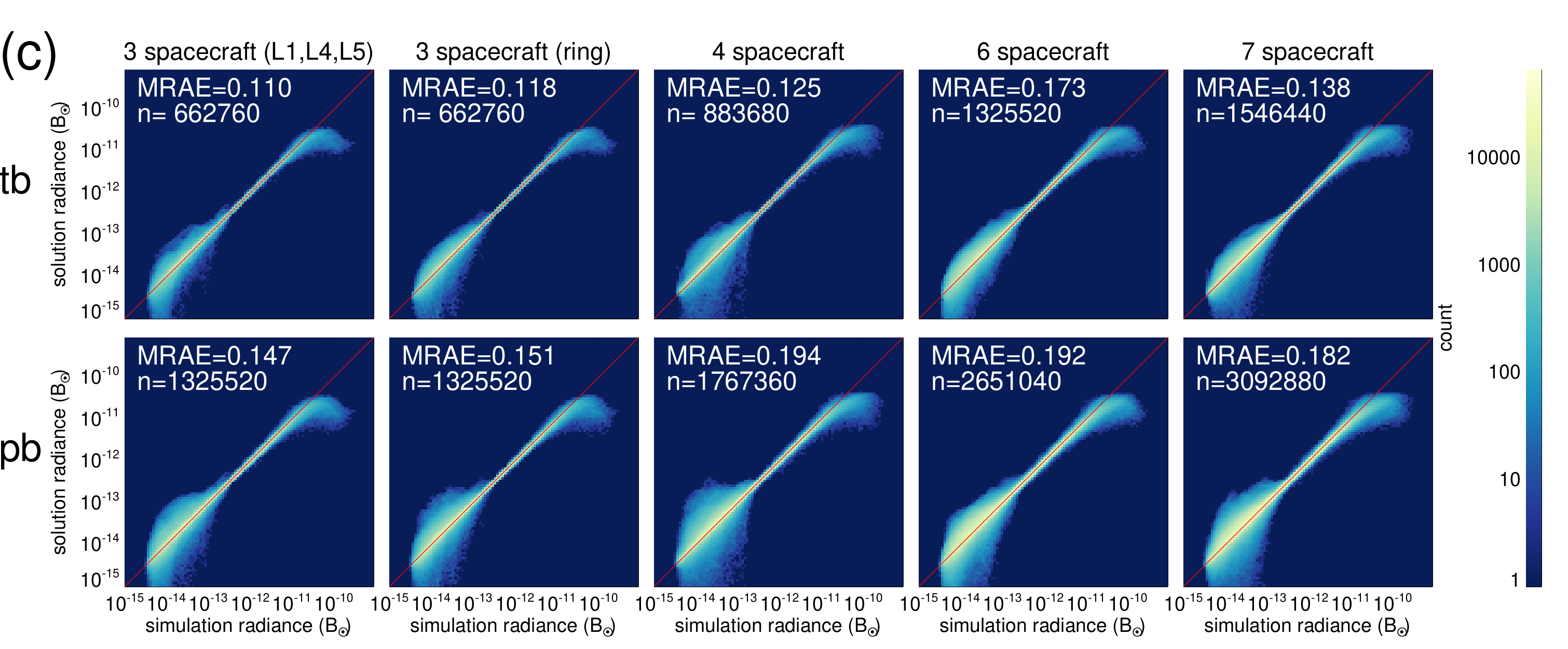}
    \caption{Two-dimensional histograms comparing radiance values in every pixel in the simulated and solution images, for each CME at every time-step. \textbf{(a)} shows results for CME1, \textbf{(b)} for CME2 and \textbf{(c)} for CME3. In each group, the top row represents solutions using tb measurements only, and the bottom row represents solutions using a combination of pb and tb. Each column represents each of the five different spacecraft configurations, labelled above the panels.The MRAE is printed in each case, as is the total number of pixels contributing to each histogram, $n$. The identity line is shown in red. The count number in each group of histograms is scaled logarithmically to emphasise low values.}
    \label{fig:radiance_scatter_hist}
\end{figure*}

As a means to assess how well the solving algorithm performs for different spacecraft configurations, and for polarimetric versus non-polarimetric reconstructions, we compare the pixel values from the simulation and solution for each CME, for every time-step. For each case study, we focus the analysis only on time-steps where the CME is visible in the processed image. This begins when the CME emerges from behind the artificial $4R_\odot$ occulter and ends when any part of the CME exceeds the $7.6^{\circ}$ FOV limit of the coronagraph. For CME1 this comprises 19 time-steps (20\,--\,38); for CME2, this comprises 24 time-steps (15\,--\,38) and for CME3 this is 21 time-steps (12\,--\,32). For each of the three CMEs, we have ten different solutions comprising the five spacecraft configurations, each of which is solved twice using either tb only, or a combination of pb and tb measurements. For a given CME, spacecraft configuration and polarisation method, we take every pixel in every image and compare the simulation to the solution. From these values we calculate the MRAE between simulation and solution. In Figure \ref{fig:radiance_scatter_hist} we bin the values in two-dimensional histograms, with a logarithmic scaling using 100 bins spanning six orders of magnitude ($10^{-15}$\,--\,$10^{-9}\,$B$_\odot$). The count number in each histogram is also scaled logarithmically in order to emphasise the lower counts that lie away from the identity line (red). CME1 is represented by the group of panels labelled \textbf{(a)}, CME2 is represented by the group of panels labelled \textbf{(b)} and CME3 by the group of panels labelled \textbf{(c)}. In each group, the top row shows the solutions using non-polarimetric (tb) measurements only and the second row shows those using polarimetric measurements (tb and pb). Each of the five columns represents one of the five spacecraft configurations, which are labelled above the plots. The total number of contributing pixels, $n$, is printed on each plot. The MRAE for all cases is consistently very good, ranging from 0.037 (CME1; 3 spacecraft ring, non-polarimetric reconstruction) to 0.194 (CME3; 4 spacecraft, polarimetric reconstruction). For all three CME case studies and polarisation methods, we find that the MRAE is greater for the 6- and 7-spacecraft configurations, than the 3- or 4-spacecraft configurations and we find that, generally, the MRAE increases when using polarimetric reconstructions over non-polarimetric reconstruction for a given configuration. We expect that this results from the image processing methods that we apply before solving the inverse equation, in particular the background subtraction. These factors have the effect of damaging the true photometry of the original images and therefore introduce inaccurate information to the solving algorithm, which prevents it from converging precisely. When we combine multiple viewpoints, or multiple polarisation measurements, we are introducing more conflicting information into the solving algorithm, therefore preventing its ability to converge and resulting in greater MRAE values between the simulated and solution radiance values. When considering the application of the method to real observations, this implies that the accuracy of image processing techniques will have a significant impact on the precision of the results, particularly as the number of observing spacecraft increases. In all cases we observe a greater spread away from the identity line that occurs at the very lowest radiance values. This is attributed to the fact that the solving algorithm is tuned to disfavour radiance values that are low in the original MAS images, based on the weighting factor in Equation \eqref{eq:weight}. This effect is least prominent in the case of CME1, and more apparent for CME2 and, particularly, for CME3. However, we reiterate that the counts in each distribution are plotted on a logarithmic scale in order to emphasise poor values, and that these features are not significant. Overall, we conclude that the solving algorithm performs well in all cases.

\subsection{Fidelity of Density Reconstructions}

\begin{figure*}
    \centering
    \includegraphics[width=0.49\textwidth]{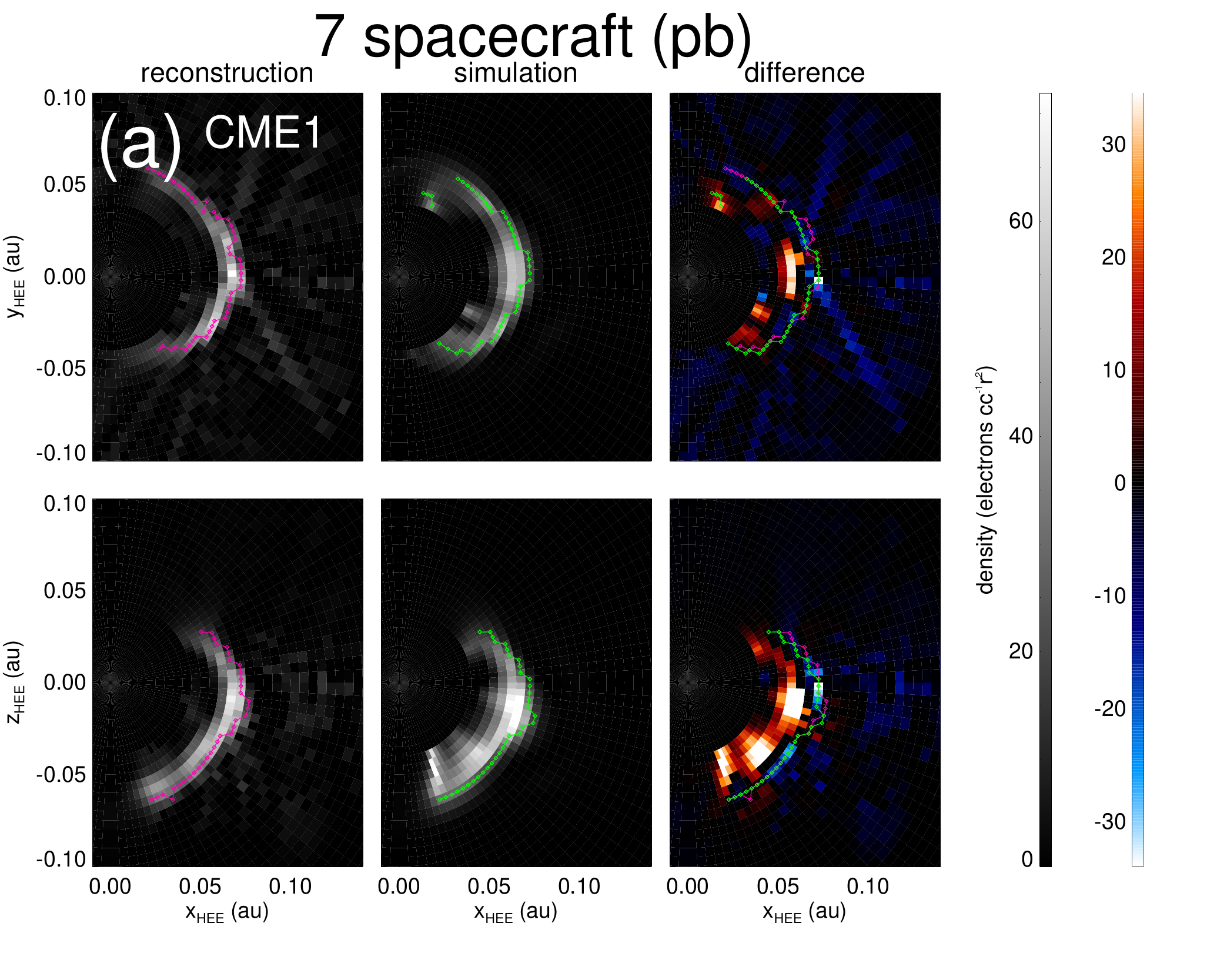}
    \includegraphics[width=0.49\textwidth]{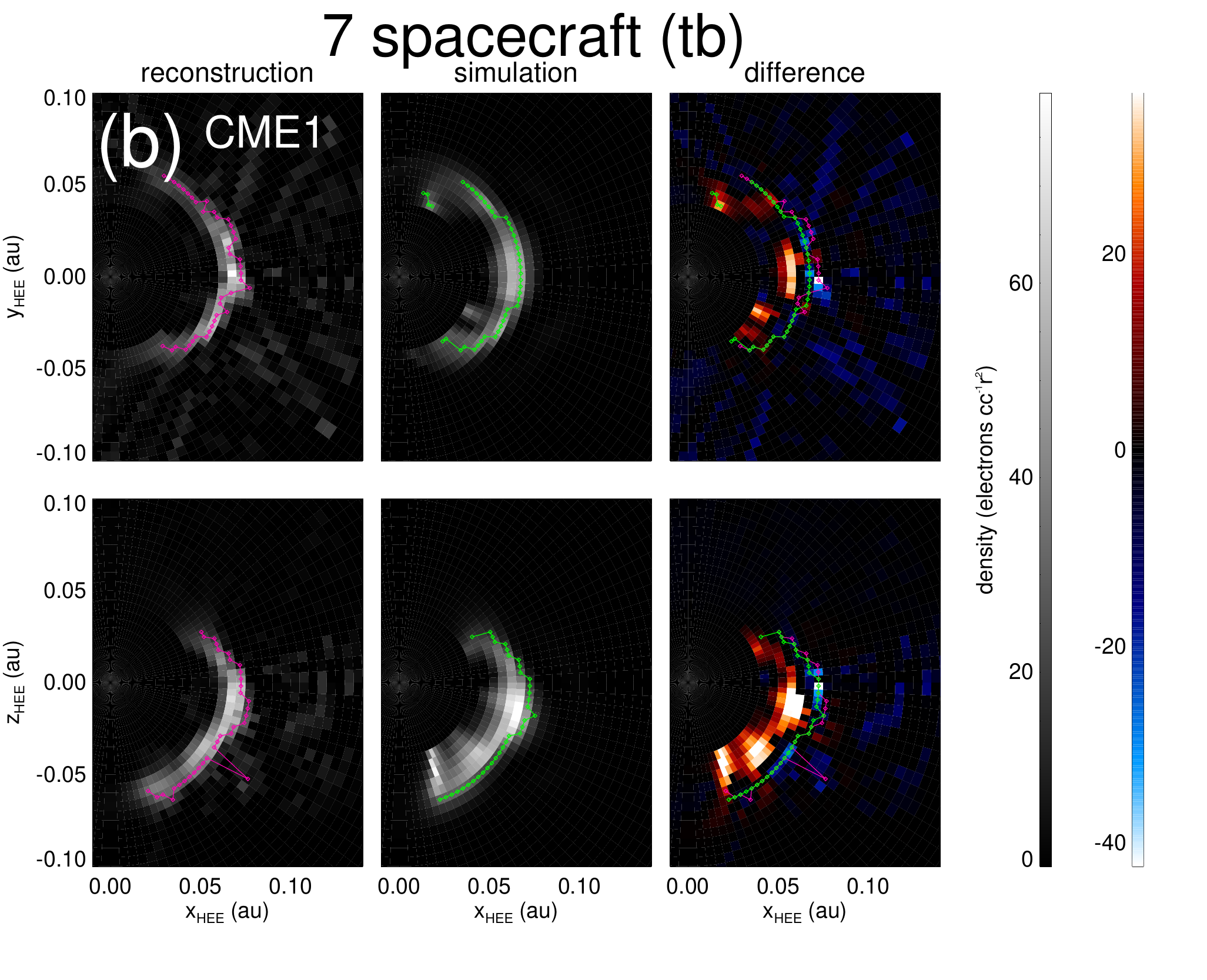}
    \includegraphics[width=0.49\textwidth]{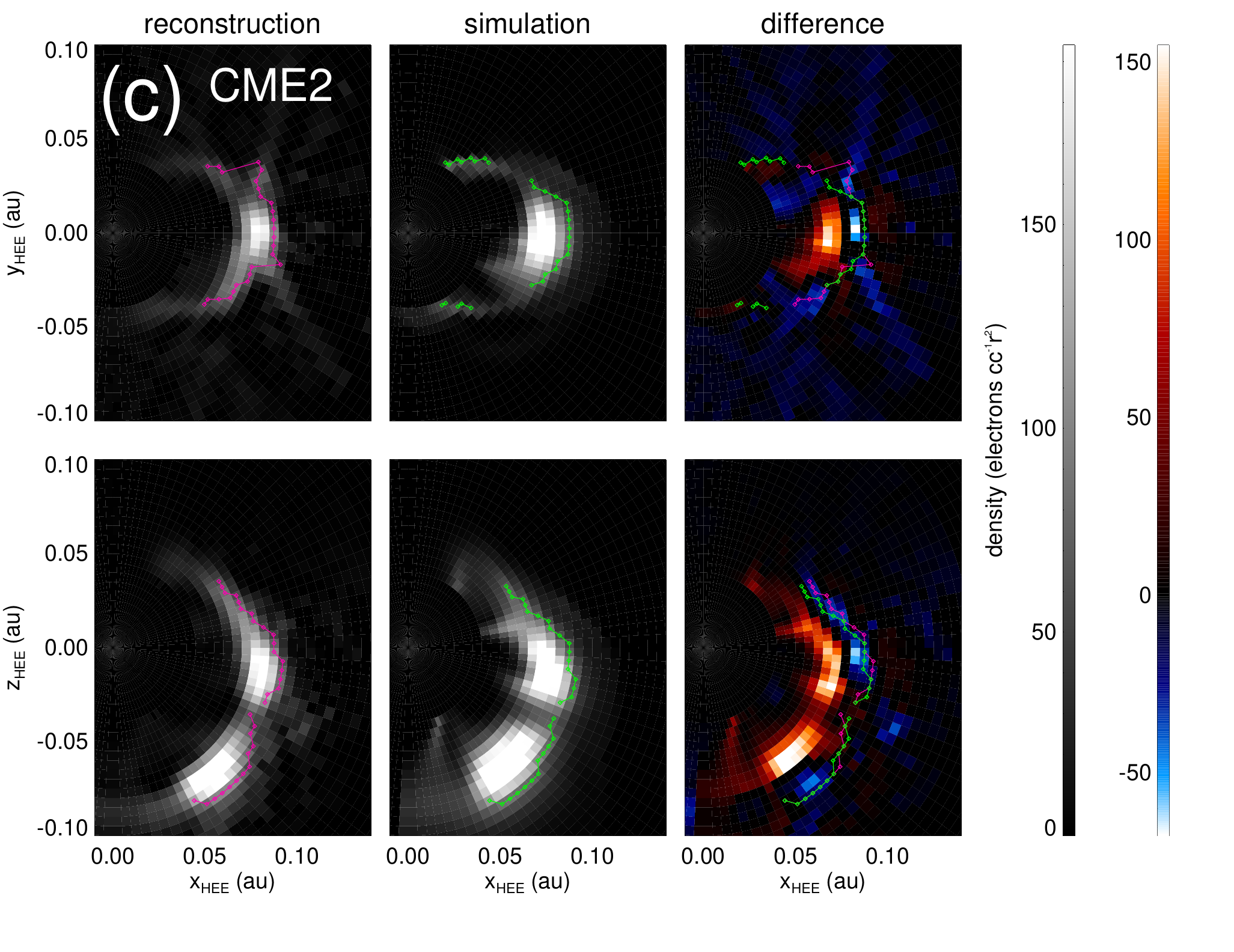}
    \includegraphics[width=0.49\textwidth]{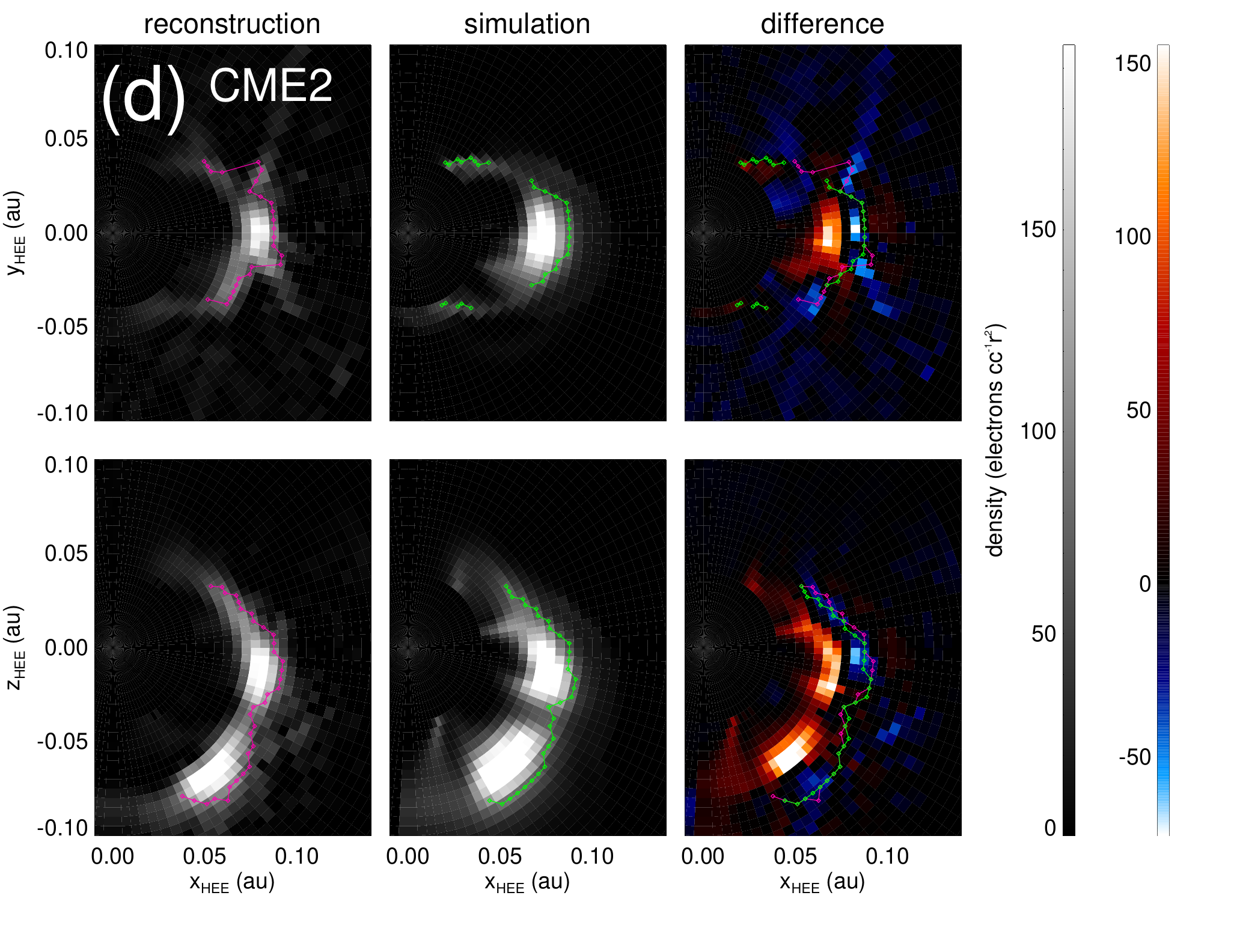}
    \includegraphics[width=0.49\textwidth]{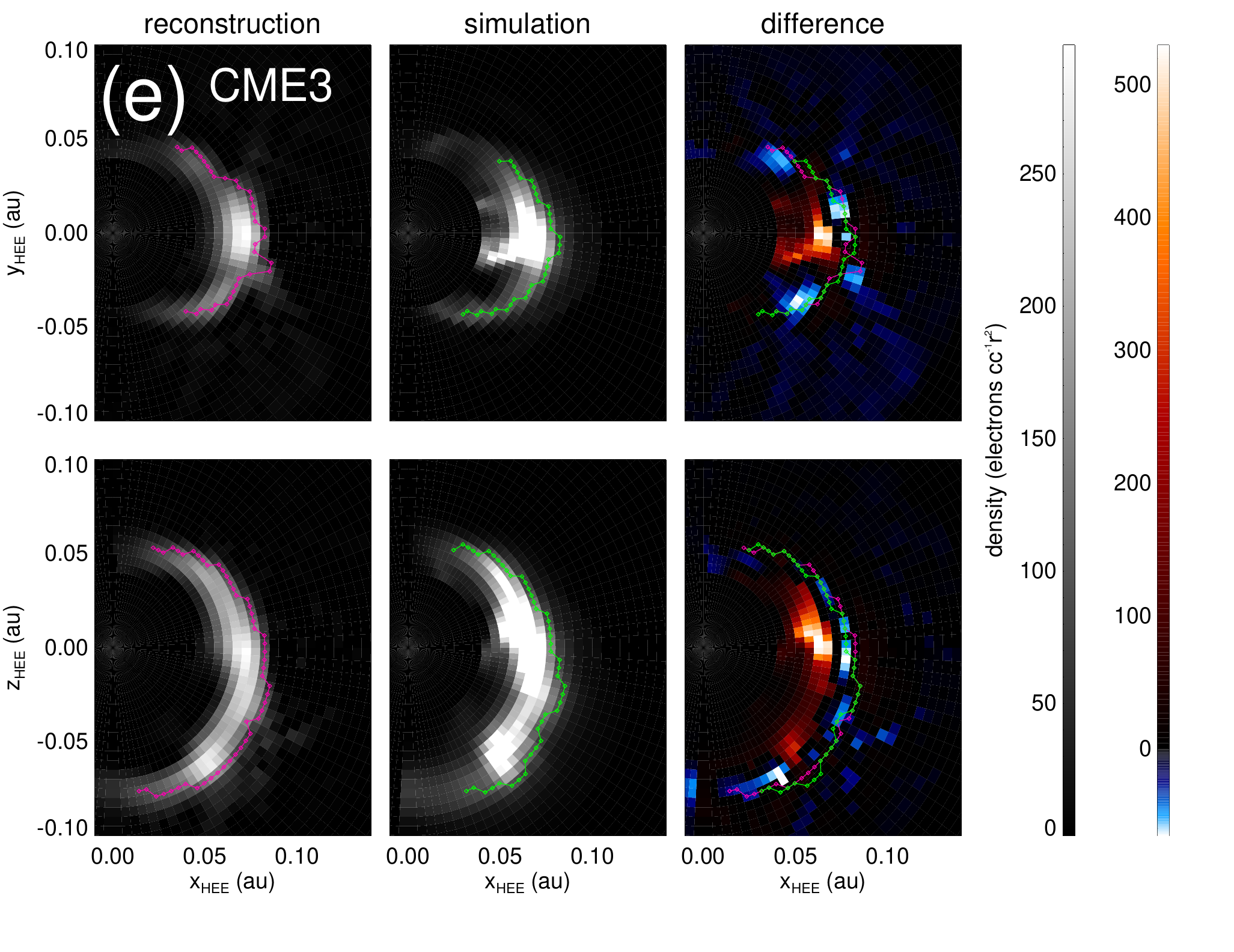}
    \includegraphics[width=0.49\textwidth]{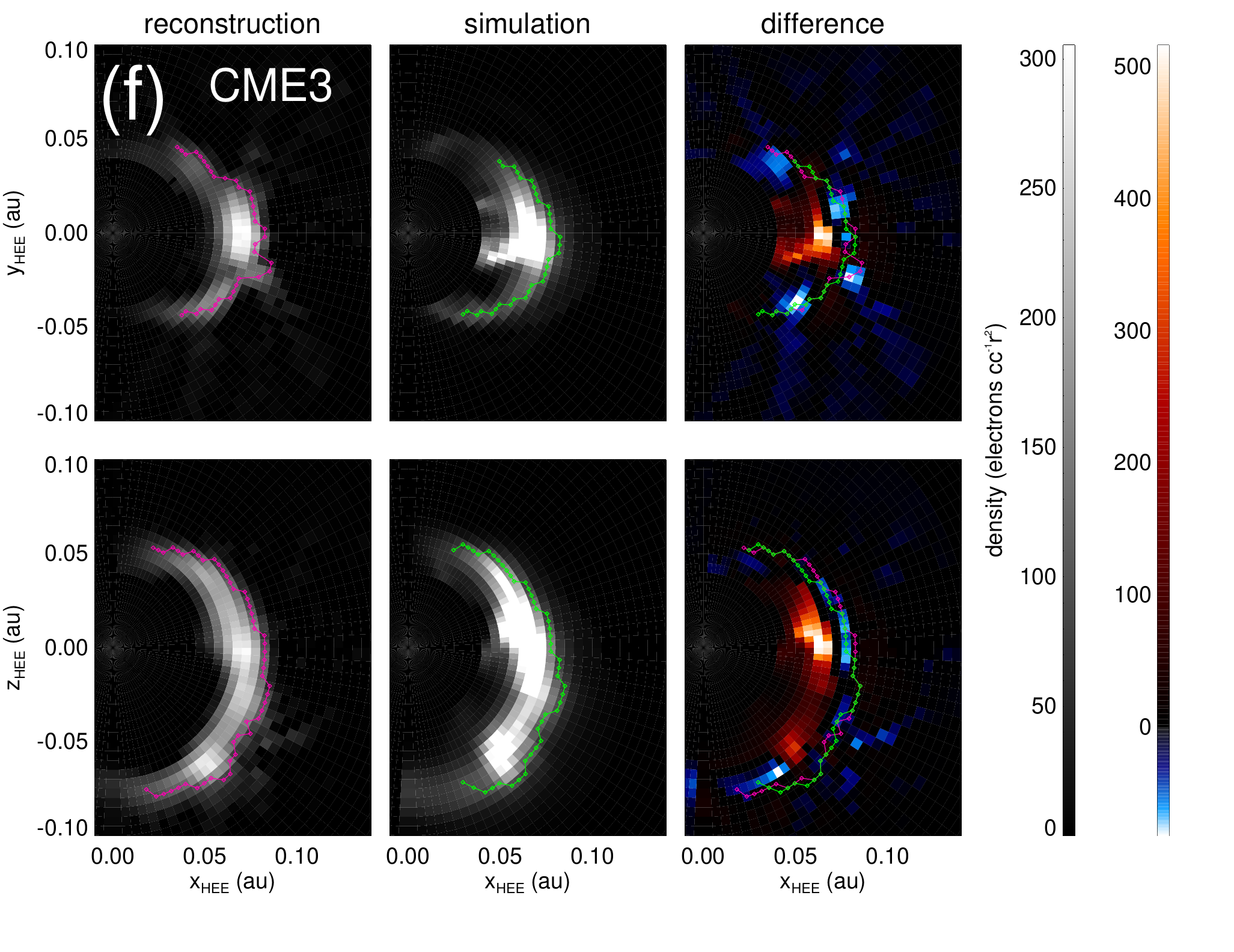}
    \caption{2D cross-sections taken from 3D density distributions. Panels \textbf{a}, \textbf{c} and \textbf{e} represent CME1 ($t=30$), CME2 ($t=28$) and CME3 ($t=20$), respectively, based on 7-spacecraft polarimetric reconstructions. Panels \textbf{b}, \textbf{d} and \textbf{f} show the equivalent non-polarimetric reconstructions. In each panel, the left hand column shows results from the tomographic inversion and the middle column shows the re-binned, background-subtracted MAS density. The right column shows the difference between the two, where positive (red) indicates greater MAS density values and negative (blue) corresponds to values from the the tomographic inversion that are greater. The top rows show the ecliptic plane and the bottom rows show the meridional plane, where the Earth lies on the $x$-axis. In each case, the location of the CME front is over-plotted in pink for the tomographic reconstructions, and green for the MAS densities.}
    \label{fig:compare_densities}
\end{figure*}

\noindent
Figure \ref{fig:compare_densities} shows examples of 2D slices in the ecliptic and meridional planes, taken from the 3D tomographic inversions using 7-spacecraft reconstructions. Panels a, c, and e correspond to CME1, CME2, and CME3 polarimetric reconstructions, respectively. Panels b, d and f show the equivalent non-polarimetric reconstructions. The reader is strongly encouraged to consult the supplementary material, available online, to view 3D visualisations of the reconstructions for all events. In each case, the re-binned, background-subtracted MAS densities are shown for comparison and the difference between the tomographic inversion and the original simulations is also shown. In all cases, we see that the general morphology of the CME front is reproduced well in the tomographic reconstructions; however, a significant amount of the internal structure is missing. We expect that this results from our image processing methods, where too much CME internal structure is removed from the images, which in turn causes the internal CME structure to be lost in the reconstruction. Applying the inversion method accurately to real observations will require robust image processing \citep[e.g.][]{2013bDeForest,2014Howard,2017DeForest}. This effect is most significant for the slowest event, CME1, where the CME front is resolved very well, but almost no internal structure is present in the reconstruction. We note that CME1 contains less internal density than the other two events (see Figure \ref{fig:mhd_sims}). The example cases in Figure \ref{fig:compare_densities} show that the reconstructed density in the meridional cross-section is more true to the simulation than in the ecliptic cross-sections. This is because it is easier to constrain the structure in the meridional plane, due to projection effects when measuring longitudinal structure from the ecliptic \citep[e.g.][]{2019Kilpua}. For all three events, the ecliptic cross-sections show LOS smearing, which is the result of performing the inversion with too few observation angles \citep[e.g.][]{1994Davila,2020aBarnes} with which to accurately constrain the density. This suggests that even seven spacecraft are likely too few to fully reconstruct 3D CME density structure. Indeed, in classical tomography methods, such as medical imagining \citep{2018Maier}, hundreds of observation angles are typically used, as is the case in established coronal rotational tomography methods \citep{2019Morgan}. However, in all three CME case studies presented here, the results of tomographic inversion can be used to locate the position of the CME front well, which is addressed in detail in Section \ref{subsec:front}.

\begin{figure*}
    \centering
    \includegraphics[width=0.49\textwidth]{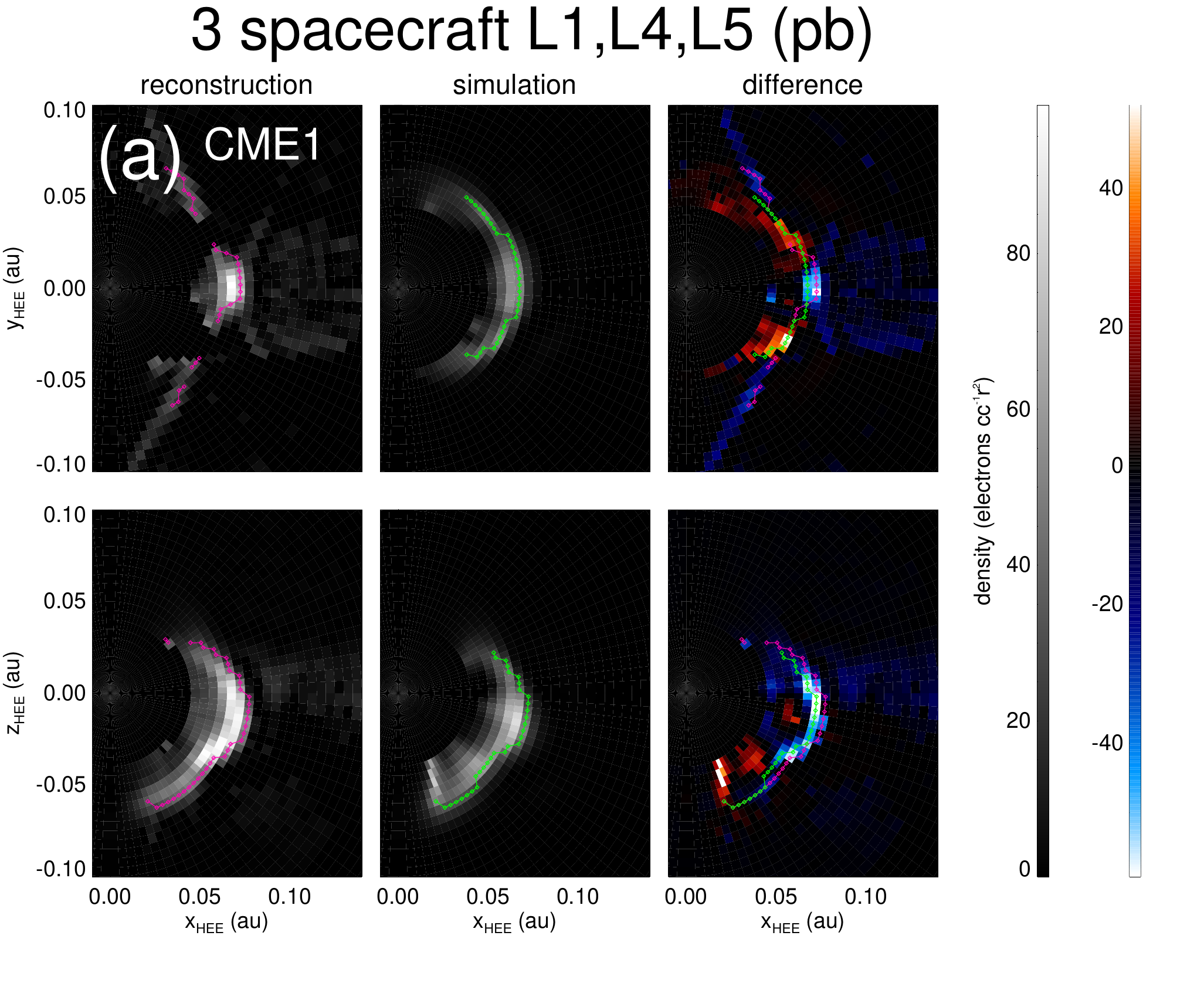}
    \includegraphics[width=0.49\textwidth]{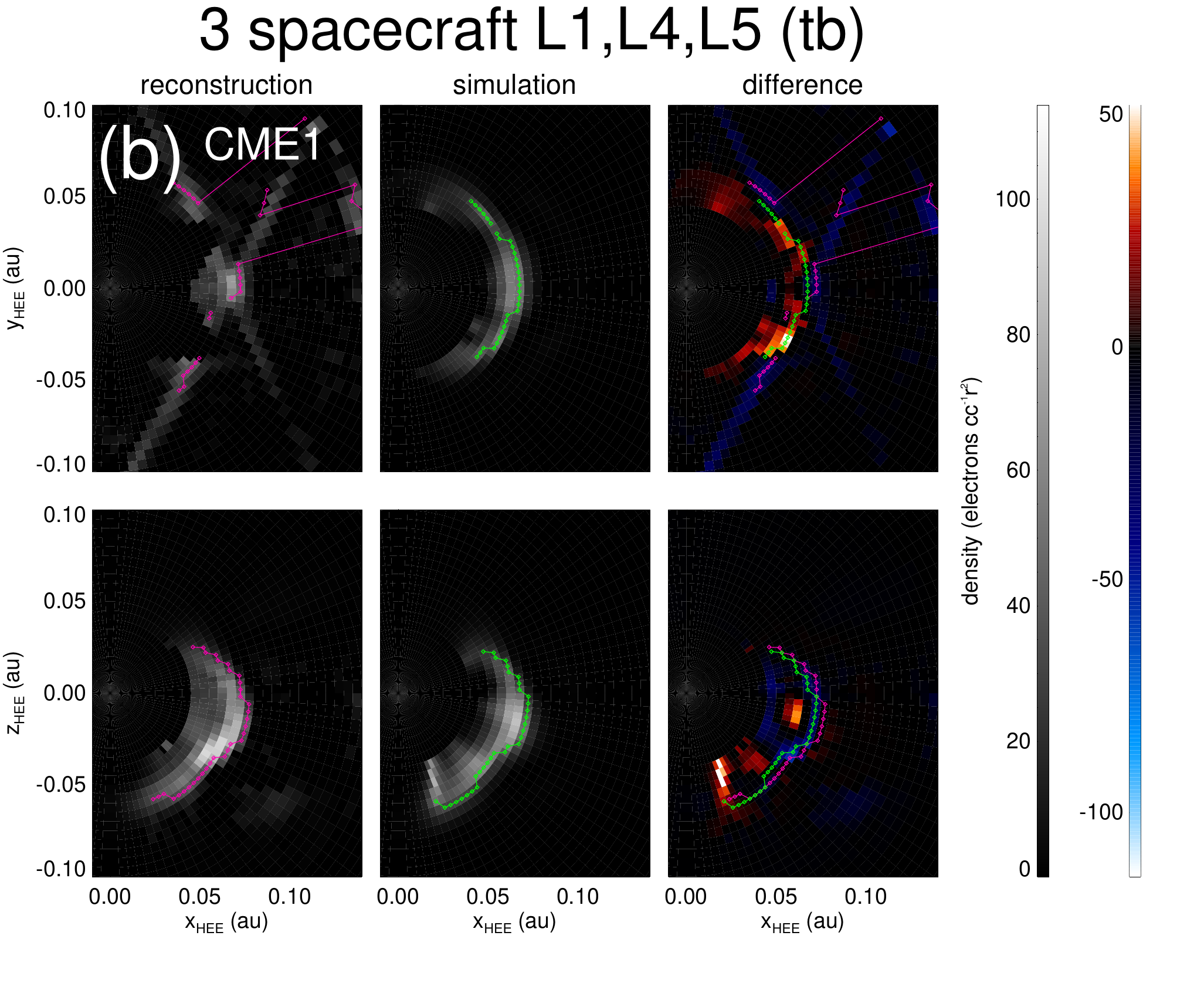}
    \includegraphics[width=0.49\textwidth]{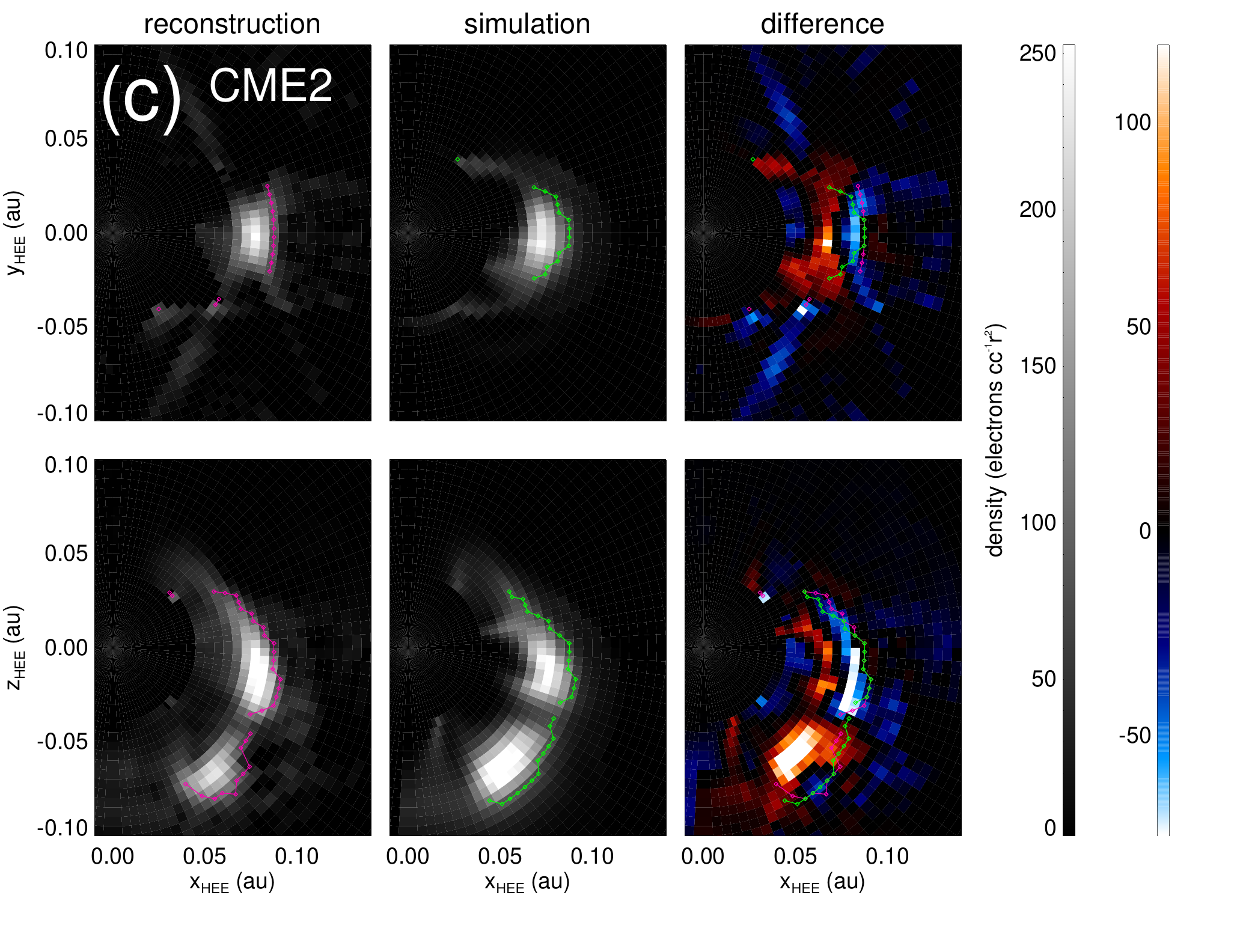}
    \includegraphics[width=0.49\textwidth]{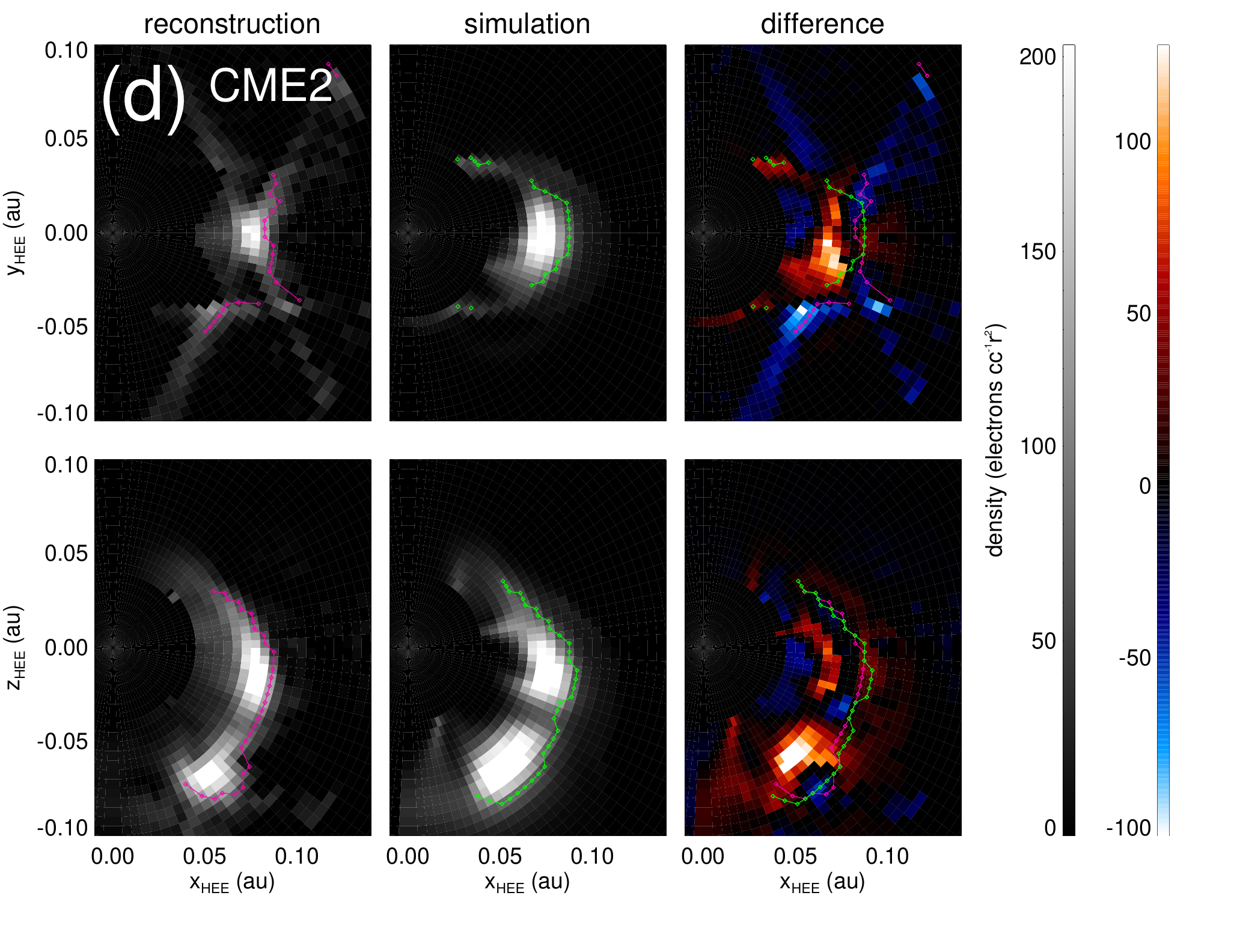}
    \includegraphics[width=0.49\textwidth]{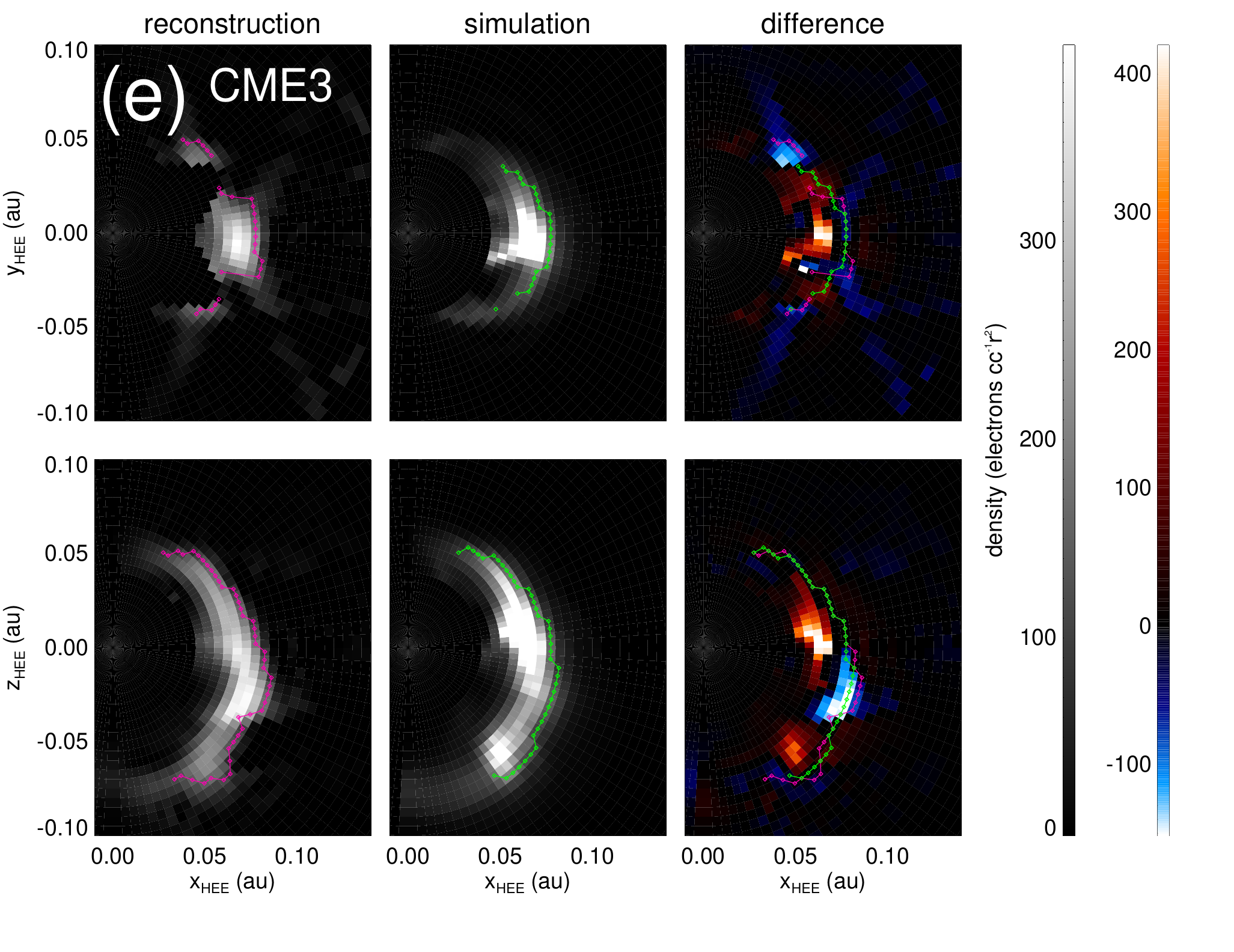}
    \includegraphics[width=0.49\textwidth]{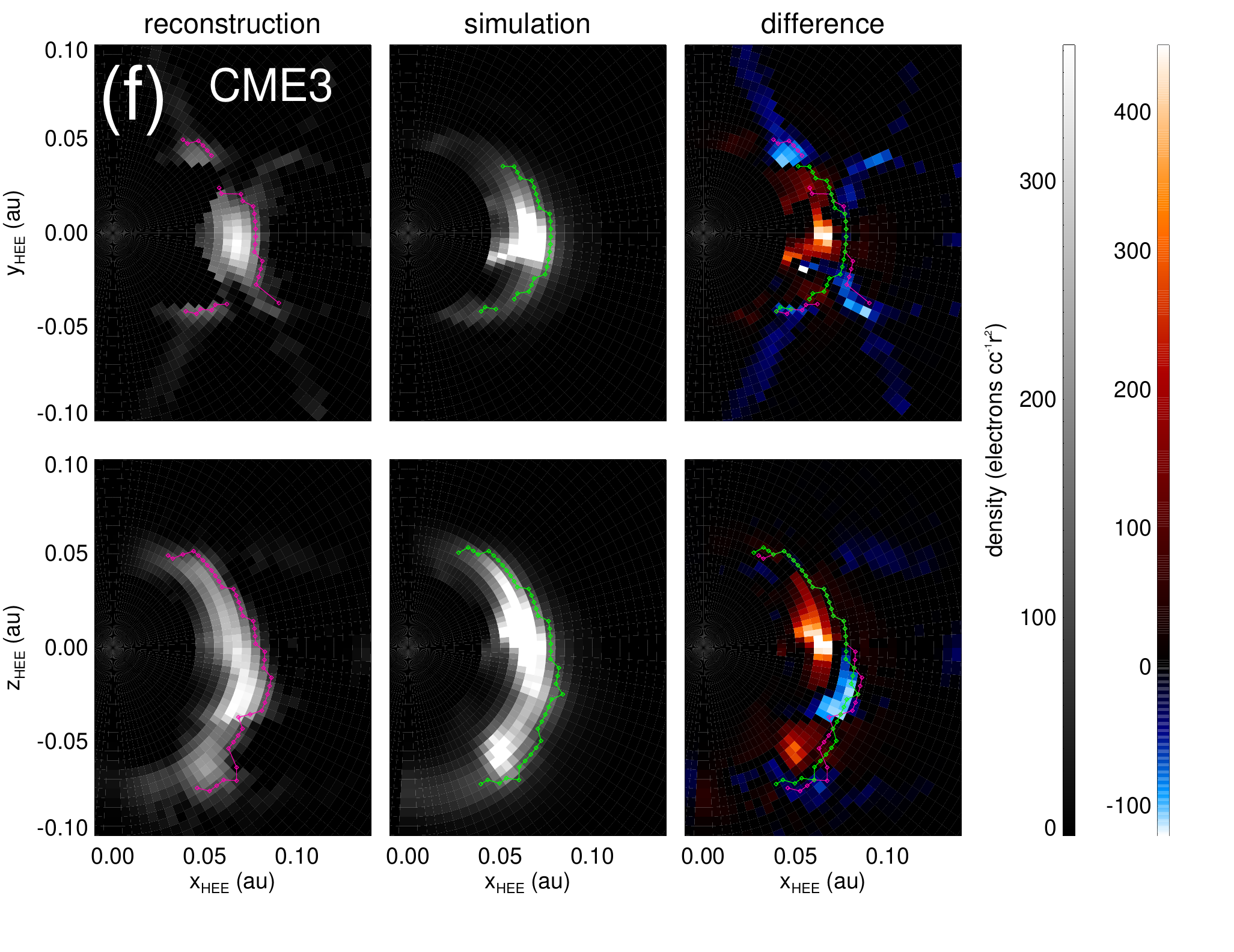}
    \caption{The same format as Figure \ref{fig:compare_densities}, showing density reconstructions using three spacecraft at L1, L4 and L5.}
    \label{fig:compare_densities2}
\end{figure*}

Figure \ref{fig:compare_densities2} shows more examples of 2D slices in the ecliptic and meridional planes, this time based on the tomographic inversions using three spacecraft at L1, L4 and L5. Again, panels a, c, and e correspond to CME1 ($t=30$), CME2 ($t=28$), and CME3 ($t=20$) polarimetric reconstructions, respectively, and panels b, d and f show the non-polarimetric reconstructions. The morphology of the CME is reasonably well reproduced in all cases, particularly in the meridional plane. However, much more LOS smearing is apparent due to the reduced number of observing spacecraft, which is particularly visible in the ecliptic cross-sections for CMEs 1 and 2. There is a more apparent difference between the polarimetric and non-polarimetric reconstructions than is apparent in the results derived using 7-spacecraft. In particular the CME front is less well defined for the non-polarimetric reconstructions, most noticeably in the case of CMEs 1 and 2. Using just three spacecraft, the solution volume is restricted due to the stipulation that we only solve for grid cells that are observed by at least two spacecraft. This results in regions of the CME at $\pm30^\circ$\ longitude near the Sun that are missing from the reconstructions, which are visible in the ecliptic cross-sections for all three CMEs. Including an out-of-ecliptic observer at P1 serves to reduce this issue significantly.

\begin{figure*}
    \centering
    \includegraphics[width=0.95\textwidth]{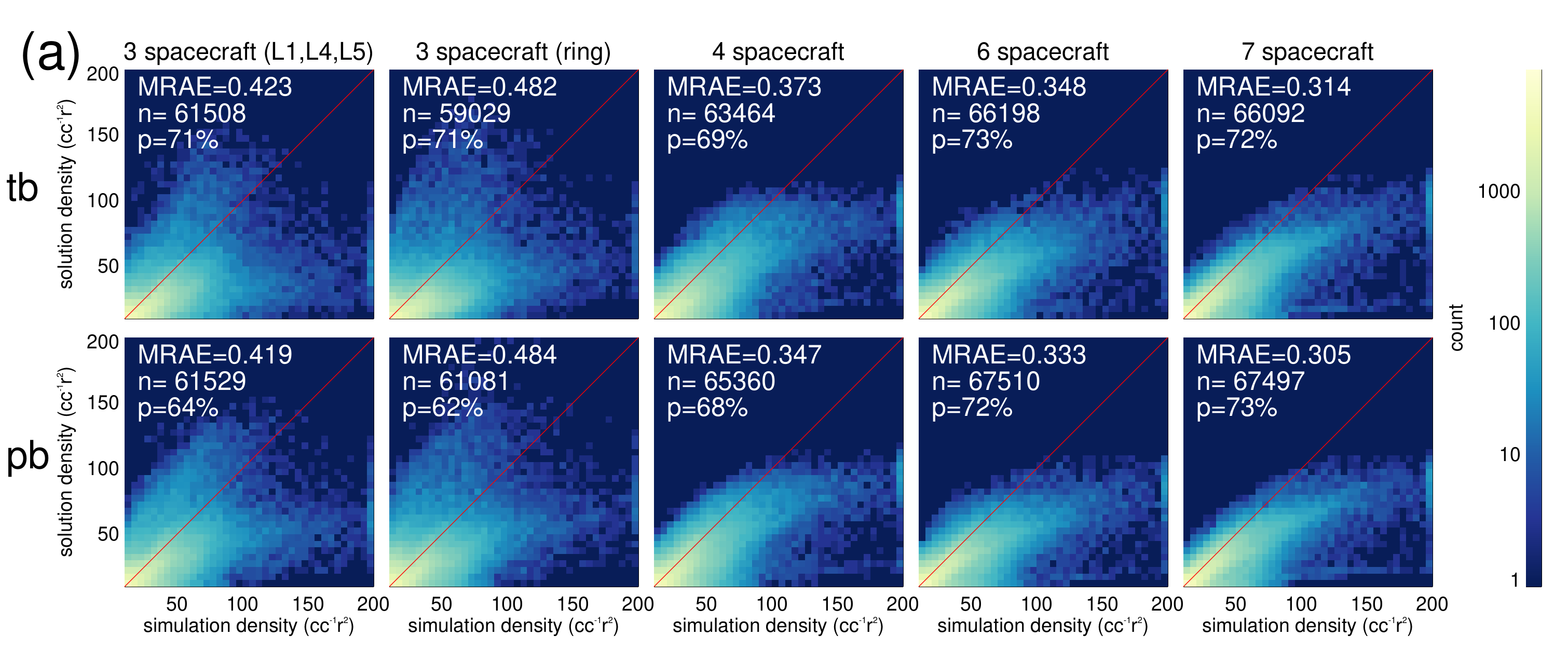}
    \includegraphics[width=0.95\textwidth]{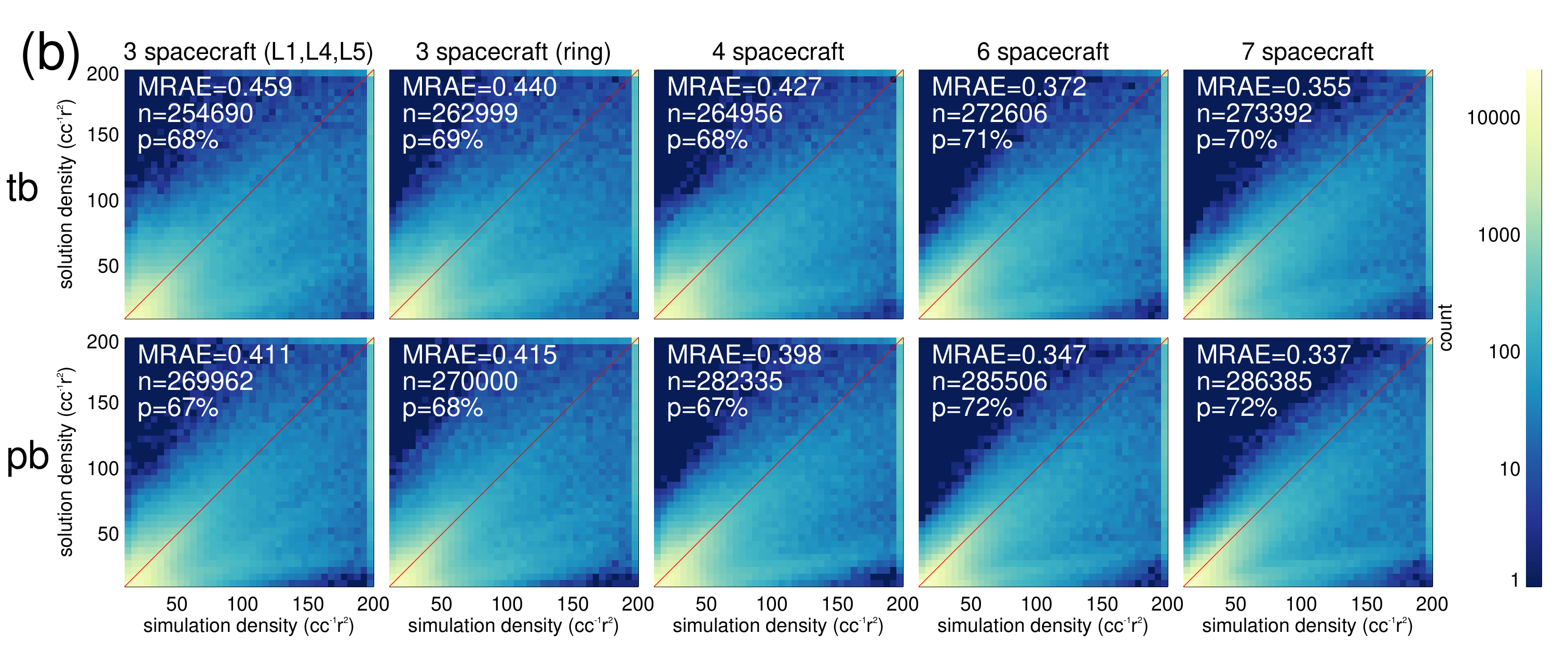}
    \includegraphics[width=0.95\textwidth]{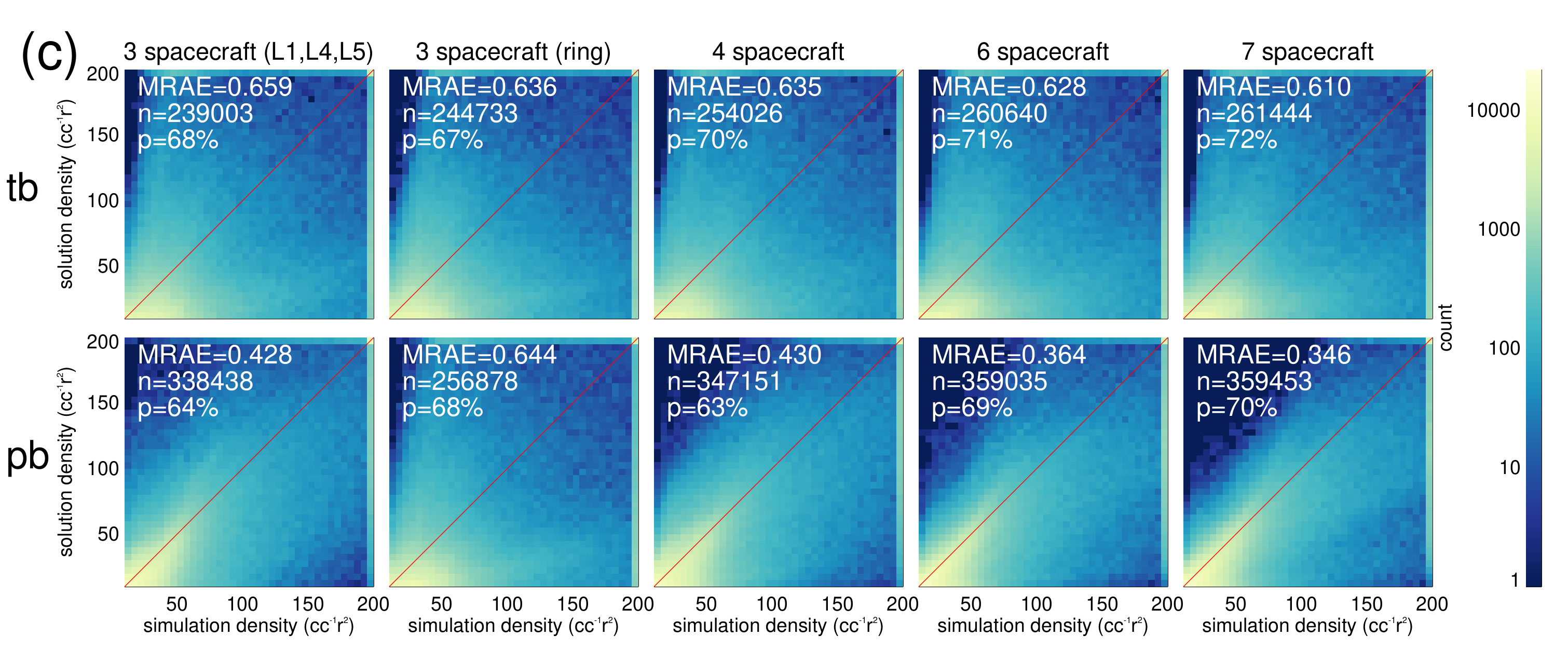}
    \caption{Two-dimensional histograms showing a comparison between density derived from tomographic reconstruction to the re-binned MAS, background-subtracted MAS densities. Panels \textbf{a}, \textbf{b} and \textbf{c} represent CME1, CME2 and CME3, respectively. In each panel, the top row uses only tb measurements in the solution and the bottom row uses both tb and pb. Each column represents different each of the five spacecraft configurations. Each bin width is 5 electrons\,$cc^{-1}r^2$ wide and only values above 10 electrons\,$cc^{-1}r^2$ are included as a means to exclude background values. $n$ shows the number of grid cells used to calculate each histogram, MRAE shows the mean relative absolute error between the simulation and solution and $p$ represents the percentage of values that lie below the identity line (red). $p$ therefore measures the rate at which the tomography underestimates density values.}
    \label{fig:density_scatter_hist}
\end{figure*}

Figure \ref{fig:density_scatter_hist} shows a comparison of all density values from tomographic inversion compared against re-binned densities from the MAS simulation. Panels a, b, and c show results for CME1, CME2, and CME3, respectively. In each panel, the top row shows results using tb only and the bottom row shows polarimetric reconstructions using both pb and tb. Each column represents one of the five spacecraft configurations, labelled above each plot. We include only grid cells where the density exceeds 10 electrons$\,cc^{-1}r^2$ as a means to focus on densities that lie significantly above the background level. Significant differences in the general distributions are seen in the case of CME1 compared to the other two events. CME1 is the slowest event, and so the background subtraction method has the effect of removing more of the CME structure from the images. This in turn causes more internal CME density to be lost from the tomographic reconstructions, which is apparent from the much lower number of counts, $n$, included in the histograms for CME1 because only grid cells over $10$ electrons\,$cc^{-1}\,r^2$ are included. Despite the apparent differences between the distributions for CME1 and the other two events, this event results in lowest MRAE values overall for when using 7-spacecraft polarimetric reconstructions. We note that the counts in Figure \ref{fig:density_scatter_hist} are plotted on a logarithmic scale to emphasise low values and the apparent differences in the shape of the distributions between CME1 and the other two events are unremarkable.  

In the case of every CME, for both polarimetric and non-polarimetric reconstructions, we find that the MRAE decreases as the number of observing spacecraft is increased, with the exception of the polarimetric reconstruction for CME3, where the 4-spacecraft configuration performs slightly worse than the 3-spacecraft at L1, L4, L5. In the case of most spacecraft configurations, for all three events, we also find that the polarimetric reconstructions result in a lower MRAE value when compared to the non-polarimetric reconstructions, with the exception of the 3-spacecraft ring for CME1 and CME3. For each of the three events the lowest MRAE is seen when applying polarimetric reconstructions to the 7-spacecraft configuration; 0.305 (CME1), 0.337 (CME2), 0.346 (CME3). Conversely, the largest MRAE is seen in non-polarimetric reconstrcutions using one of the 3-spacecraft configurations; 0.459 for CME2 and 0.659 for CME3. For CME1, the worst case is the polarimetric reconstuction using the 3-spacecraft ring configuration (0.484). After each iteration, the solving algorithm returns some negative values, which are reset to the corresponding initial guess (1 electron $cc^{-1}r^2$), which causes the tail that can be seen towards the bottom edge in many of the distributions. The solving algorithm consistently underestimates density values, which results from the background subtraction method removing much of the CME internal structure. The average number of density values that are underestimated are 71\% (tb) and 68 \% (pb) for CME1, 55\% (tb) and 69\% (pb) for CME2 and 70\% (tb) and 67 \% (pb) for CME3.

\begin{table}
    \centering
    \begin{tabular}{c|c|c|c|c|c}
        event    & 3-sc (L1, L4, L5) & 3-sc (ring) & 4-sc & 6-sc & 7-sc \\
         \toprule
         CME1 &  0.9\% & $-$0.4\% &  7.0\% &  4.3\% &  2.9\% \\
         CME2 & 10.5\% &  5.7\% &  6.8\% &  6.7\% &  5.1\% \\
         CME3 & 35.1\% & $-$1.3\% & 32.3\% & 42.0\% & 43.3\% \\
    \end{tabular}
    \caption{Relative reduction in MRAE between simulated and reconstructed density when using polarimetric reconstructions over non-polarimetric reconstructions, for each spacecraft configuration and for each CME case study.}
    \label{tab:pb_ratio}
\end{table}

Table \ref{tab:pb_ratio} shows the relative improvement in MRAE values when using polarimetric over non-polarimetric reconstructions for each spacecraft configuration and for each CME. In each case, we consistently find that polarimetric reconstructions give better results than those using only tb images, with the exception of the 3-spacecraft ring for CME1 and CME3. We observe no significant trend that this improvement correlates to the number of spacecraft used in the reconstruction. However, for CME3, the polarimetric reconstructions appear to provide more of a benefit when a larger number of spacecraft are included in the analysis. This is because the non-polarimetric reconstructions for this event are poor, however, the reason for this is unclear.

\subsection{CME Localisation} \label{subsec:front}
As a means to quantify the ability of our tomography method to resolve the CME structure, we focus on the CME \editone{front}, which is the feature that is typically tracked in CME propagation studies \citep[e.g.][]{2020Barnard,2022Temmer}. Here, the \editone{CME front} is calculated from the 3D density distributions, both for the tomographic inversions and the re-binned, background subtracted MAS densities, using the method described in Section \ref{subsec:validation}. Figure \ref{fig:compare_densities} shows examples of CME fronts calculated using this method applied to the 7-spacecraft reconstructions and Figure \ref{fig:compare_densities2} shows examples using three spacecraft at L1, L4 and L5. \editone{In these figures, the CME front in the simulation and reconstruction often match exactly, or to within one grid cell, particularly in the meridional plane. This is likely the result of resampling the MAS densities on to a lower resolution grid. We expect that a more accurate assessment of the errors in the location of the CME front could be achieved by running the tomography at higher resolution.}

\begin{figure}
    \centering
    \includegraphics[width=0.45\textwidth]{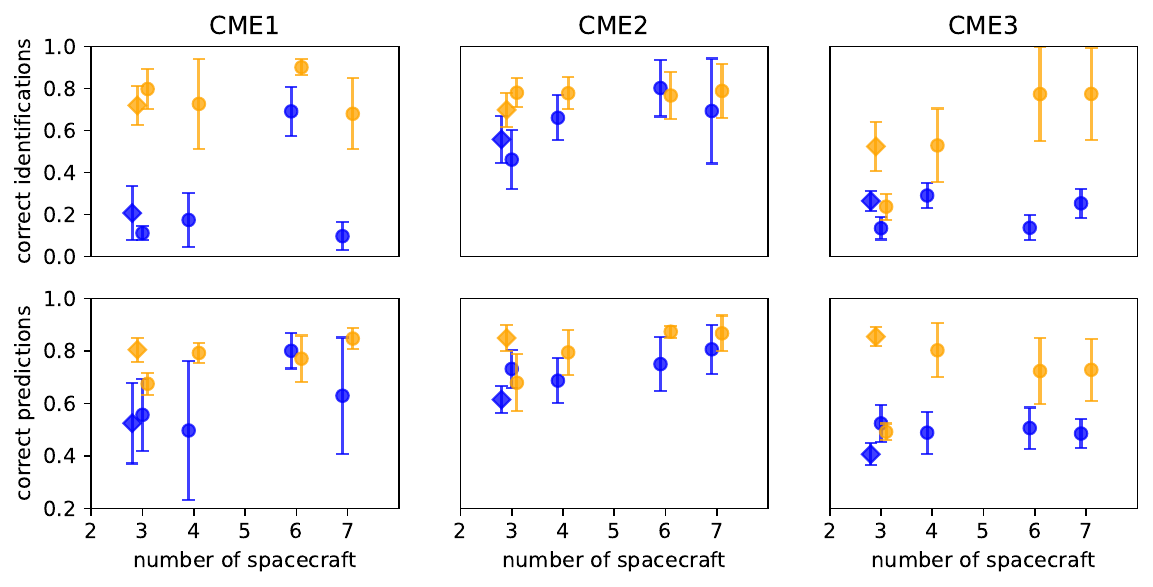}
    \caption{Plots showing the ability of tomography to locate the CME front. \emph{Correct identifications} measure the number of instances that a CME front is present at a given latitude/longitude in the MAS density, which is also identified by the tomography, i.e., the fraction of correct MAS values that are identified in the reconstruction. \emph{Correct predictions} are the fraction of instances that a CME front is identified in the tomographic reconstruction at a given latitude/longitude that also corresponds to a CME front that is present from the MAS simulation, i.e. the fraction of predictions from the tomography that are correct. Values shown are averages and standard deviations calculated over the time-steps during which the CME is observed. Blue values represent non-polarimetric reconstructions and yellow values represent polarimetric reconstructions. In the cases where three spacecraft are used, the L1, L4, L5 configuration is represented by a diamond, and the 3-spacecraft ring is represented by a circle.}
    \label{fig:front_hits}
\end{figure}

Figure \ref{fig:front_hits} shows the ability of the tomographic inversion to correctly identify the CME front for each of the three events. The top row of plots, labelled \emph{correct identifications}, represents the fraction of instances where a CME front is identified in the MAS density distribution, at a given latitude and longitude, that is also correctly predicted in the tomography. The bottom row, labelled \emph{correct predictions}, shows the fraction of instances where a CME front is predicted in the tomography that also corresponds to a front identified in the MAS density. We find that the correct identifications from polarimetric reconstructions are good for all events; with an average of $(76\pm8)\%$, $(76\pm3)\%$ and $(57\pm20)\%$ for CME1, CME2, and CME3, respectively, regardless of the number of observers. The corresponding values for the non-polarimetric reconstructions are $(26\pm22)\%$, $(63\pm12)\%$, $(22\pm7)\%$; showing that the polarimetric reconstructions are appreciably more accurate at determining the existence of the CME front. Across the three events, we do not observe a trend that increasing the number of observers improves the correct identification value. Using just three spacecraft at L1, L4 and L5; polarimetric reconstructions are able to correctly identify the CME front with an accuracy of $(72\pm9)\%$, $(70\pm8)\%$ and $(52\pm12)\%$ for CME1, CME2 and CME3, respectively. This level of accuracy cannot be achieved using any number of spacecraft in the non-polarimetric reconstructions, with the exception of CME2, when six spacecraft are used $(80\pm14)\%$.

The rate of correct predictions varies between $41\%$ and $87\%$ for all cases, regardless of the number of spacecraft, showing that the presence of a CME front is consistently over-predicted using the tomographic inversions. There is no significant trend in the improvement of the correct predictions versus number of observing spacecraft. For all spacecraft combinations, the rate of correct predictions averages $(60\pm11)\%$ (CME1), $(72\pm6)\%$ (CME2) and $(48\pm4)\%$ (CME3) using non-polarimetric reconstructions and $(78\pm6)\%$ (CME1), $(81\pm7)\%$ (CME2) and $(72\pm12)\%$ (CME3) using polarimetric resconstructions. This shows that the non-polarimetric method over-predicts the existence of CME structure, whilst the polarimetric method is more reliable in every case. We observe no trend in the correct predictions based upon the number of spacecraft.

\begin{figure*}
    \centering
    \includegraphics[width=0.99\textwidth]{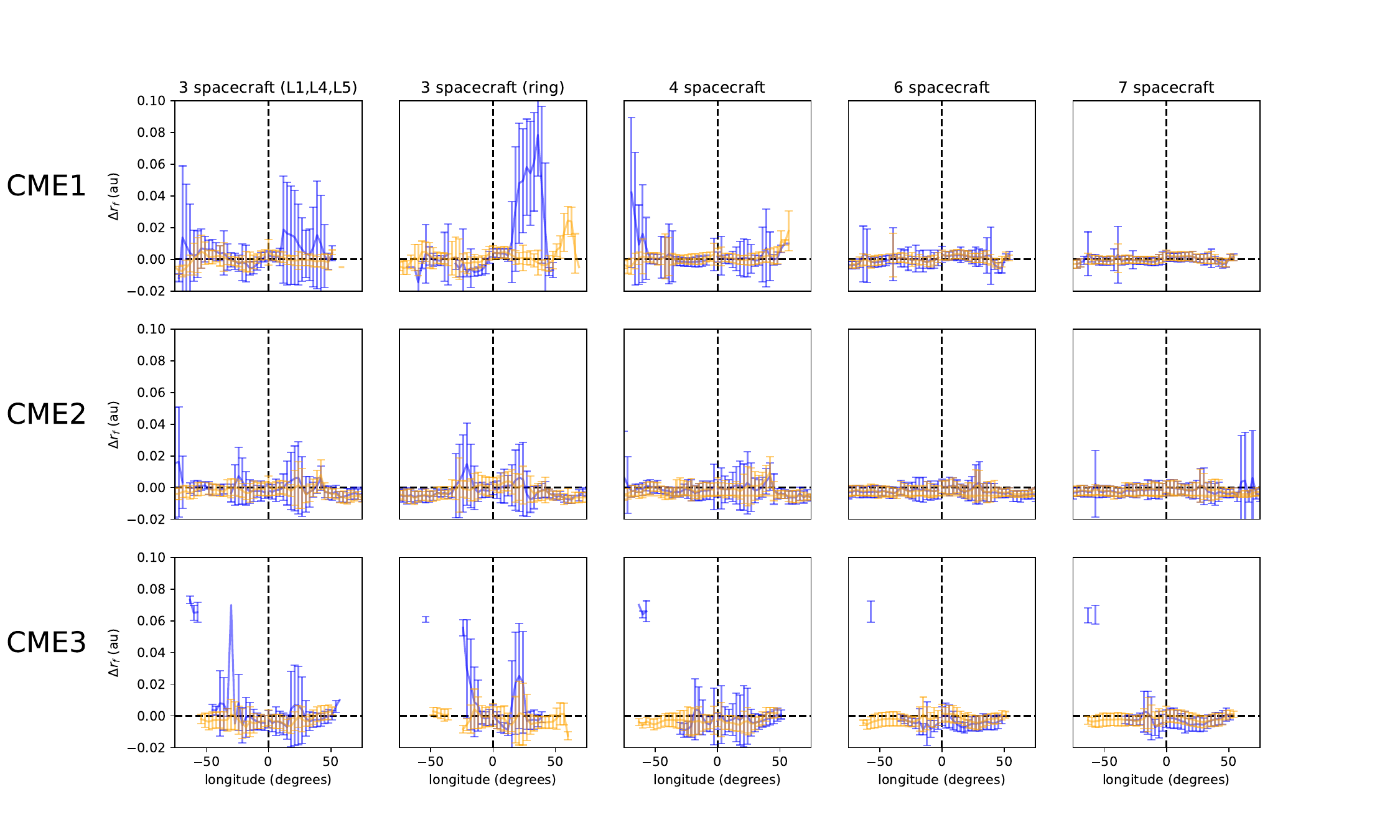}
    \caption{Difference in radial distance of the CME front between tomographic reconstructions and re-binned background-subtracted MAS densities, $\Delta r_f$, averaged over grid-cell longitude from all time-steps. The top, middle and bottom rows represent CME1, CME2 and CME3, respectively. Columns represent each of the five spacecraft combinations, labelled above each. Blue values represent non-polarimetric reconstructions and yellow values represent polarimetric reconstructions.}
    \label{fig:front_lon}
\end{figure*}

Figure \ref{fig:front_lon} shows the difference in radial location of the CME front, $\Delta r_f$, averaged as a function of grid-cell longitude, between the tomographic reconstructions and re-binned, background-subtracted MAS densities. This difference is calculated for instances where the CME front is identified in both the tomographic inversion and in the MAS simulations, i.e. the correct identifications. We see that the location of the CME front is identified well, typically within 0.02\,au, in all cases. In all cases we see that the polarimetric reconstructions produce lower values of $\Delta r_f$ overall than the non-polarimetric results. For all CMEs we observe a trend that increasing the number of spacecraft results in lower values of $\Delta r_f$ for both polarimetric and non-polarimetric methods. CME reconstructions using fewer -- 3 or 4 -- observing spacecraft exhibit larger and more variable values of $\Delta r_f$ than when using 6 or 7 observers, and these discrepancies are more pronounced in the non-polarimetric reconstructions. This indicates a greater ability to accurately reconstruct the CME front as the number of observing spacecraft is increased. For all three CME cases studies, we see that the position of the CME front across all longitudes can be determined to a high level of precision using 4-spacecraft polarimetric reconstructions. The 4-spacecraft non-polarimetric reconstructions locate the CME front reasonably well, but to a lesser degree. In the case of CME3, non-polarimetric reconstructions fail to identify CME limb structure corresponding to values away from zero longitude when compared to the polarimetric reconstructions.

\begin{figure*}
    \centering
    \includegraphics[width=0.99\textwidth]{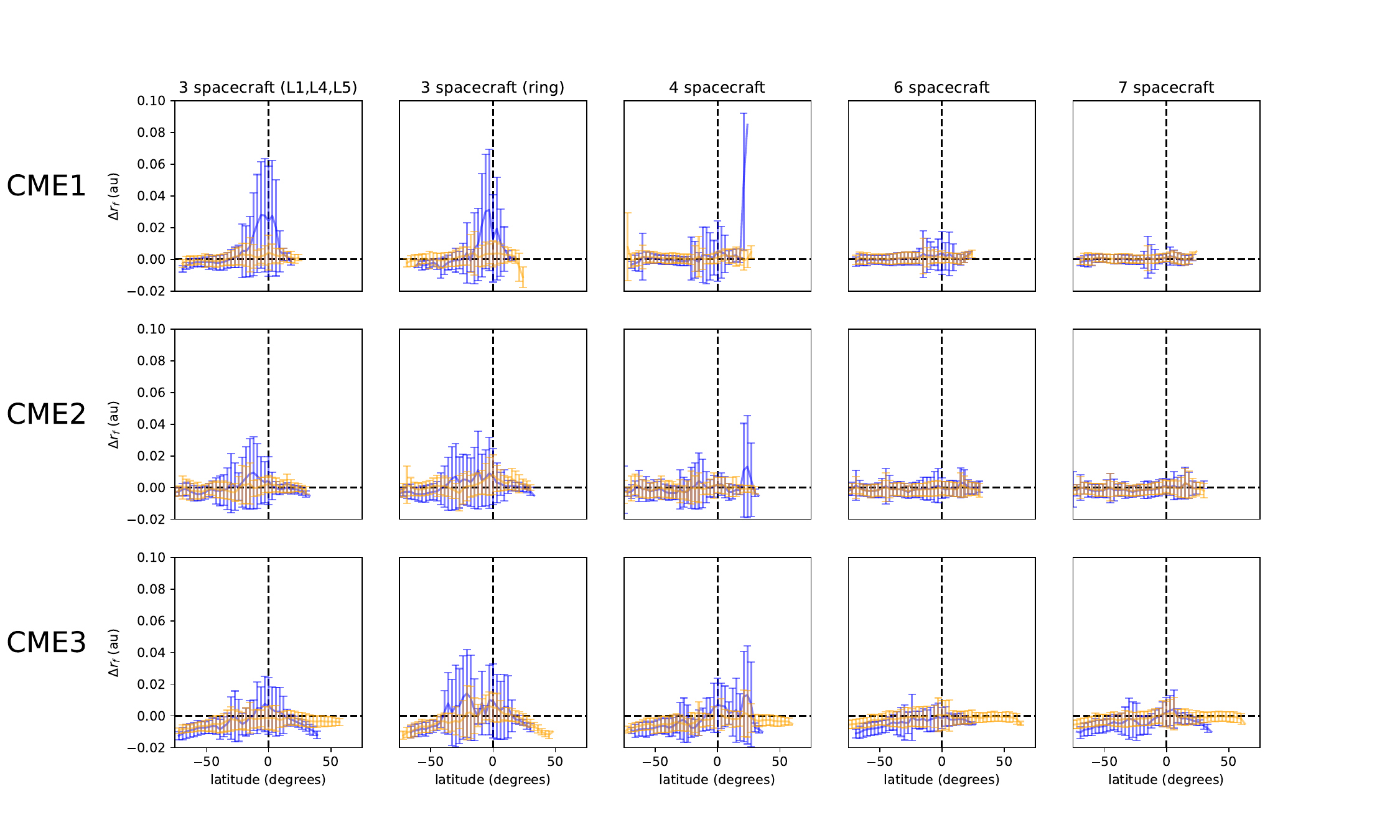}
    \caption{The same format as Figure \ref{fig:front_lon}, where values are averaged as a function of grid-cell latitude, rather than longitude.}
    \label{fig:front_lat}
\end{figure*}

The variation in $\Delta r_f$ over latitude for each CME and each spacecraft combination is shown in Figure \ref{fig:front_lat}. Again, the polarimetric reconstructions are consistently better and the number of observing spacecraft is seen to improve the accuracy of the CME front localisation. Polarimetric reconstructions using four or more spacecraft are able to constrain the CME front to a high degree of precision, whilst at least six are required using non-polarimetric methods. When using six or seven spacecraft, there is little difference in the accuracy of the polarimetric and non-polarimetric reconstructions. Again, the full extent of CME3 is much better constrained using the polarimetric method, whilst a significant amount of northward CME structure (above $40^\circ$) is missing in the non-polarimetric reconstructions.

\begin{figure}
    \centering
    \includegraphics[width=0.45\textwidth]{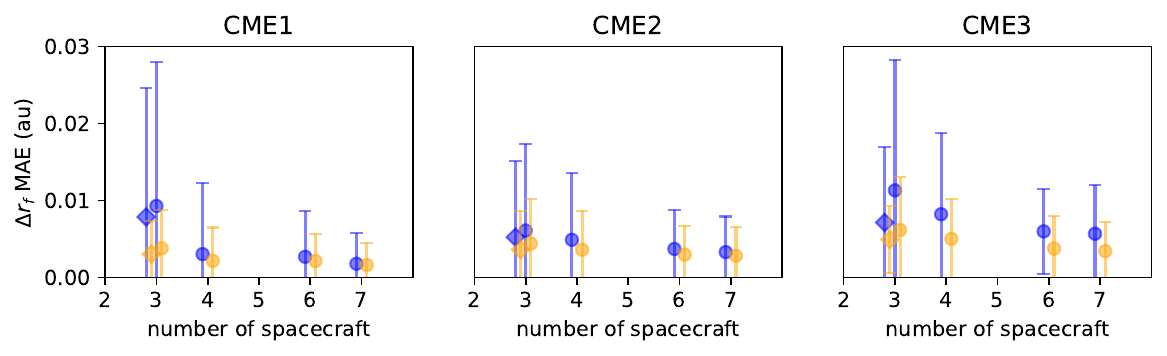}
    \caption{MAE in $\Delta r_f$ averaged over all longitudes, latitudes and time-steps as a function of spacecraft number and polarisation method. Blue represents non-polarimetric reconstructions and yellow represents polarimetric reconstructions. In the cases where three spacecraft are used, the L1, L4, L5 configuration is represented by a diamond, and the 3-spacecraft ring is represented by a circle.}
    \label{fig:front_scs}
\end{figure}

Figure \ref{fig:front_scs} shows the MAE in values of $\Delta r_f$, averaged over all latitudes, longitudes and time-steps where the front is correctly identified in the tomographic reconstructions. For every spacecraft configuration, in all three events, it is clear that the MAE using polarimetric reconstuctions is lower than when using non-polarimetric reconstructions. Furthermore, the precision of each measurement is also greater when using polarimetric reconstructions. For every CME, there is a slight trend that the MAE values decrease as the number of observing spacecraft increases. A significant increase in the precision of the MAE values is observed in the non-polarimetric reconstructions as the number of spacecraft increases, and the same trend is seen in the polarimetric results, but to a lesser extent. Using just three spacecraft at L1, L4 and L5, the polarimetric reconstructions produce $\Delta r_f$ MAE values of $0.003\pm0.004$\,au, $0.004\pm0.005$\,au and $0.005\pm0.004$\,au for CME1, CME2 and CME3, respectively. In the case of CME1 and CME2, seven spacecraft are required to reach the same level of precision ($0.002\pm0.004$\,au and $0.003\pm0.005$\,au) using non-polarimetric reconstructions and, for CME3, this precision is not matched by any of the non-polarimetric reconstructions. It is noteworthy that the L1,L4,L5 configuration consistently performs slightly better than the 3-spacecraft ring, despite both methods using the same number of observers.

\section{Conclusions}
We investigate the effectiveness of a discrete tomography method, which is used to reconstruct 3D CME density structure from the inversion of synthetic coronagraph images from multiple spacecraft. We apply the analysis to three separate simulated CME case studies as a means to test the method. The method is applied using five different spacecraft combinations ranging from three to seven observers, and the reconstructions are performed twice in each instance using polarimetric and non-polarimetric reconstructions.

We find that the solving algorithm works well in all cases, with a MRAE between the simulated and reconstructed images ranging from 0.037 in the best case (CME1; 3-spacecraft ring, non-polarimetric reconstruction) to 0.194 in the worst case (CME3; 4-spacecraft, polarimetric reconstruction). In the cases where pb measurements are included, our method has the capability to reproduce the polarimetric properties of the simulated images well. We employ a simple background subtraction method to isolate the CME signal and find that it removes too much internal structure from the CMEs, particularly in the case of the slowest ($\sim1000\,km\,s^{-1}$) event, CME1. This processing impairs the photometry of the original images and MRAE values between simulated and reconstructed radiance values increase somewhat as more spacecraft are included in the analysis and more conflicting information is introduced to the solving algorithm.

Over-subtraction of background means that the reconstructed density measurements are generally lower than in the simulations, where we underestimate between 62\% and 73\% of the density values greater than 10 electrons $cc^{-1}r^2$. \editone{Established methods used to calculate CME mass from real white light images \citep[e.g.][]{2009colaninno,2017deKoning} depend on measurements of image brightness and are likely to suffer similarly from background subtraction effects. However, in our reconstructions, the overall CME morphology appears to be unaffected by this over-subtraction.} Accurate photometric inversion methods will require rigorous image processing \citep[e.g.][]{2013bDeForest,2014Howard,2017DeForest} in order to reconstruct the internal density structure of CMEs. Despite this, our inversion method is able to reproduce density structure well, with the MRAE ranging from 0.659 (CME3; 3-spacecraft (L1, L4, L5), non-polarimetric reconstruction) at worst, to 0.305 (CME1; 7-spacecraft, polarimetric reconstruction) in the best case. In all cases, increasing the number of observing spacecraft serves to reduce the MRAE between simulated and reconstructed density. The largest MRAE is seen in when using three spacecraft for each of the three CMEs, regardless of whether polarimetric or non-polarimetric reconstructions are used. Likewise, the lowest MRAE is seen when using 7 spacecraft in all cases. We also find that polarimetric reconstructions, compared to non-polarimetric reconstructions, improve the MRAE between simulated and reconstructed density in all but two cases, with the relative improvement ranging from -1.3\% (CME3; 3-spacecraft ring) to 43.3\% (CME1; 7-spacecraft). There is no significant trend that this improvement changes as the number of spacecraft is increased.

Three of the five spacecraft configurations employed in this study include spacecraft in the ecliptic plane only, whilst the other two include a polar spacecraft $60^\circ$ above the ecliptic. We do not find a conclusive improvement when including the polar spacecraft, rather we find that the total number of observing spacecraft is the most significant factor in increasing the fidelity of the density reconstructions. We expect that a greater number of spacecraft combinations would need to be tested in order to thoroughly study the benefits afforded by out-of-ecliptic observations. However, we find that the volume of space over which the tomographic inversion may be applied is restricted when using just three spacecraft and that the inclusion of an out-of-ecliptic observer would be necessary in order to maximise coverage close to the Sun, due to the way in which the FOVs of observing spacecraft overlap each other.  

Methods applied to locate the presence of the CME front within density reconstructions show a significant improvement when using polarimetric over non-polarimetric methods The average rate at which the presence of a CME front is correctly identified is $(76\pm8)\%$ (CME1), $(76\pm3)\%$ (CME2) and $(57\pm20)\%$ (CME3) using polarimetric reconstructions and $(26\pm22)\%$ (CME1), $(63\pm12)\%$ (CME2) and $(22\pm7)\%$ (CME3) using non-polarimetric reconstructions. The rate of correct predictions is also significantly higher using polarimetric reconstructions. The rate of correct identifications is $(60\pm11)\%$ (CME1), $(72\pm6)\%$ (CME2) and $(48\pm4)\%$ (CME3) using non-polarimetric reconstructions and $(78\pm6)\%$ (CME1), $(81\pm7)\%$ (CME2) and $(72\pm12)\%$ (CME3). Therefore all methods tend to over-predict the existence of CME structure, although this effect is much less severe when using polarimetric reconstructions. Polarimetric reconstructions using three spacecraft at L1, L4 and L5 identify the CME front with an accuracy of $(72\pm9)\%$, $(70\pm8)\%$ and $(52\pm12)\%$ for CME1, CME2 and CME3, respectively. Non-polarimetric reconstructions are not able to reproduce this level of accuracy, regardless of the number of spacecraft used, with the exception of CME2 using six spacecraft. In cases where the CME front is correctly identified, its radial position can typically be located to within 0.02\,au. For all three CME case studies, the level of precision increases slightly as more spacecraft are used in the polarimetric reconstructions and it increases greatly if six or more are used in the non-polarimetric reconstructions.

We emphasise that we are able to determine a significant amount of information about CME structure, even when using just three observing spacecraft and employing polarimetric reconstructions. Furthermore, this method is improved significantly when the number of observing spacecraft is increased. A number of studies \citep[e.g.,][]{2010Mierla,2016Jang,2023Verbeke} have highlighted the errors inherent in using traditional forward-modelling to constrain CME parameters and it is our intention to compare the reliability of forward and inverse modelling applied to these same synthetic events in future work. In this study, we apply a simple method to identify the CME front, but we are still able to demonstrate the capability of inverse modelling to accurately constrain three dimensional CME structure. This provides a potential means to infer far more information about CME morphology and propagation, such as velocity and acceleration profiles across the CME front, as well as information about CME mass and density evolution. Performing these types of measurements is beyond the scope of this study and we also aim to investigate them in further work. These considerations highlight the significant potential strengths of inverse modelling when compared to conventional forward modelling techniques. The simulated images employed in this study contain no F-corona, stars nor planets. We do, however, apply some basic processing to separate CME signal from the background and we find that this has some detrimental effects on the photometry of the images. Nevertheless, we expect that performing such analysis on real images is achievable given advanced image processing techniques that continue to be developed \citep[e.g.][]{2013bDeForest,2014Howard,2017DeForest}. Indeed, similar photometric analysis is already applied to real observations in numerous rotational tomography methods \citep[e.g.][]{2009Kramar,2010Butala,2019Morgan}.

Recent forward modelling studies \citep[e.g.,][]{2023Verbeke} have suggested that increasing the number of observing spacecraft (from two to three) offers little improvement in the accuracy of CME tracking, nor our ability to accurately forecast Earth arrivals. However, based on the findings presented here, we insist to the contrary; observations from greater numbers of spacecraft present an opportunity to employ more advanced CME tracking methods -- specifically, inverse modelling -- and will allow us to discover far more detailed information about their physical behaviour and their space weather impacts. Whilst the analysis presented in this study has yet to be achieved with existing observations, we argue that it is an opportune time to consider such methods given the current fleet of coronal and heliospheric imaging spacecraft (SOHO, STEREO, PSP, SolO, PUNCH, PROBA-3, and SWFO-L1), and upcoming missions such as Vigil. Furthermore, future mission proposals to L4 \citep{2021Posner} and polar orbits \citep{2023Hassler}, as well as multi-spacecraft polarimetric imaging missions, (e.g., MOST, \citet{2024Gopalswamy}), are likely to expand this fleet further. We hope that the results presented here may be used to better inform such mission and instrument planning in the future.

\section*{Acknowledgements}
This research was supported by the International Space Science Institute (ISSI) in Bern, Switzerland, through ISSI International Team project \#587: \emph{Tomographic Inversion of Synthetic White-Light Images: Advancing Our Understanding of CMEs in 3D}. 
\editone{The simulations analysed in this work were performed using the Expanse high-performance computer (\href{https://doi.org/10.1145/3437359.3465588}{https://doi.org/10.1145/3437359.3465588}) at the San Diego Supercomputer Center through allocation MCA03S014 from the Advanced Cyberinfrastructure Coordination Ecosystem: Services \& Support (ACCESS) programme. ACCESS is an advanced computing and data resource programme supported by the U.S.\ National Science Foundation (NSF) under the Office of Advanced Cyberinfrastructure awards \#2138259, \#2138286, \#2138307, \#2137603 and \#2138296.}
DB acknowledges support via the RAL Space In House Research programme funded by the Science and Technology Facilities Council of UK Research and Innovation (award ST/M001083/1). EP acknowledges NASA's Living With a Star (LWS; Grant 80NSSC24K1108), \editone{LWS Strategic Capabilities (LWS-SC; Grant 80NSSC22K0893), and Heliophysics Supporting Research (HSR; Grant 80NSSC25M7101) programmes}, as well as NSF's Solar, Heliospheric, and INterplanetary Environment (SHINE; Grant AGS-2301403) programme. EA acknowledges support from the Finnish Research Council (Research Fellow grant number 355659). MB is supported by ERC grant (HELIO4CAST, 10.3030/101042188), funded by the European Union. Views and opinions expressed are however those of the author(s) only and do not necessarily reflect those of the European Union or the European Research Council Executive Agency. Neither the European Union nor the granting authority can be held responsible for them. GMC acknowledges support from the Young Researchers Program (YRP), project number AVO165300016. PH acknowledges support from the Office of Naval Research. This research was funded in whole or in part by the Austrian Science Fund (FWF) [10.55776/P36093]. For open access purposes, the author has applied a CC BY public copyright license to any author-accepted manuscript version arising from this submission. We thank the developers of Python -- in particular, SciPy \citep{2020Virtanen} -- which has helped our work greatly.

\section*{Data Availability} 
The MAS model's source code is openly available at \href{https://github.com/predsci/mas}{https://github.com/predsci/mas}, and CORHEL-CME is available for Runs-on-Request at NASA's Community Coordinated Modeling Center (CCMC) via the link \href{https://ccmc.gsfc.nasa.gov/models/CORHEL-CME~1}{https://ccmc.gsfc.nasa.gov/models/CORHEL-CME\textasciitilde1}. Files containing electron density values derived from all of the tomographic reconstructions, and the re-binned values from MAS simulations, are openly available and can be accessed at \href{https://doi.org/10.5281/zenodo.18892319}{https://doi.org/10.5281/zenodo.18892319}.

\bibliographystyle{mnras}
\bibliography{refs} 








\bsp	
\label{lastpage}
\end{document}